\documentclass[12pt,preprint]{aastex}
\shorttitle{Cooling Cores}
\shortauthors{Motl, Burns, Loken, Norman \& Bryan}

\begin{document}

\title{Formation of Cool Cores in Galaxy Clusters via Hierarchical 
Mergers}
\author{Patrick M. Motl and Jack O. Burns}
\affil{Center for Astrophysics and Space Astronomy, 
University of Colorado, Boulder, CO 80309}
\author{Chris Loken}
\affil{Canadian Institute for Theoretical Astrophysics, McLennan 
Labs, University of Toronto, 60 St. George, Toronto, Ontario M5S 3H8}
\author{Michael L. Norman}
\affil{Center for Astrophysics and Space Sciences, University of 
California, San Diego, 9500 Gilman Drive, La Jolla, CA 92093}
\author{Greg Bryan}
\affil{University of Oxford, Astrophysics, Keble Road, Oxford, OX1 3RH}
\email{motl@casa.colorado.edu}
%\slugcomment{Submitted to the Astrophysical Journal}

\begin{abstract}

We present a new scenario for the formation of cool cores in rich
galaxy clusters based on results from recent
high spatial dynamic range, adaptive mesh Eulerian 
hydrodynamic simulations of large-scale structure formation.
We find that cores of cool gas, material that would
be identified as a classical cooling flow based on its X-ray luminosity
excess and temperature profile, are built from the accretion of discrete, 
stable subclusters.  Any ``cooling flow'' present is overwhelmed by the 
velocity field within the cluster - the bulk flow of gas through the
cluster typically has speeds up to about 2,000 $\mathrm{km} \;
\mathrm{s}^{-1}$ and significant rotation is frequently present in the cluster core.  
The inclusion of consistent initial 
cosmological conditions for the cluster within its surrounding 
supercluster environment is crucial when simulating the evolution of 
cool cores in rich galaxy clusters.  This new model for the hierarchical 
assembly of cool gas naturally explains the high frequency of cool 
cores in rich galaxy clusters despite the fact that a majority of these 
clusters show evidence of substructure which is believed to arise from 
recent merger activity.  Furthermore, our simulations generate complex 
cluster cores in concordance with recent X-ray observations of cool 
fronts, cool ``bullets'', and filaments in a number of galaxy clusters.  
Our simulations were computed with a coupled 
N-body, Eulerian, adaptive mesh refinement, hydrodynamics cosmology
code that properly treats the effects of shocks and radiative cooling 
by the gas.  We employ up to seven levels of refinement to attain a peak 
resolution of 15.6 h$^{-1}$ kpc within a volume 256 h$^{-1}$ Mpc on 
a side and assume a standard $\Lambda$CDM cosmology.

\end{abstract}


\section{Introduction}

The potential impact of radiative cooling by the hot intracluster medium 
(ICM) has been recognized for many years \citep{cowie77,fabian77}.
The X-ray emitting gas is trapped in approximate hydrostatic equilibrium 
in the cluster's potential well.  A ``cooling flow'' is believed to form 
as the radiating gas loses pressure support and flows inwards to higher 
density values thus accelerating the cooling rate.  In cooling flow 
clusters, typically, the gas within approximately 100 kpc of the cluster 
center has a cooling time less than the age of the cluster.  This
theoretical scenario is well known (see Fabian 1994 for a review) and, 
for example, has been examined extensively for its effect on cluster 
dominant galaxies and to determine the ultimate fate of the cool gas.

Cores of cool gas were found to be quite common in the flux-limited 
\textit{ROSAT} sample of Peres \textit{et al.} (1998), occurring in 70 to 90 
percent of their clusters.  Similarly, White \textit{et al.} (1997) 
found cool cores in $62 \pm^{12}_{15}$ \% of the 207 clusters in 
their \textit{Einstein} sample.  Though these samples are flux limited and 
will exhibit bias towards intrinsically bright core X-ray clusters, they 
nonetheless argue for cool cores being a common occurrence in 
nearby clusters.

Substructure is also a common characteristic in galaxy clusters.  For 
example, in the REFLEX+BCS cluster sample, Schuecker \textit{et al.} 
(2001) find evidence for substructure in the central 1 Mpc (for 
$H_{0} = 50 \: \mathrm{km} \: \mathrm{s}^{-1} \: \mathrm{Mpc}^{-1}$) 
in a majority ($52 \pm 7$\%) of their clusters.  The presence of 
substructure argues for accretion or merger events that occurred 
sufficiently recently to not be erased by relaxation of the cluster to 
a spherically symmetric state \citep{roettiger96}.  

The prevalence of both substructure and cool cores presents a puzzle
as one might expect mergers to disrupt the cores of cool gas present in 
one or both clusters before their merger.  Sparks (1992), however, argued 
that it was ultimately accretion events that powered the X-ray excess in 
cooling flow clusters through conductive heat transport from the hot ICM 
into accreted cores of cool gas.  While this scenario was later shown by 
\citet{fabian94} to be untenable on theoretical grounds, this does not 
negate the observational arguments that Sparks presented for a correlation 
between cooling flow and merger signatures in clusters.

Early theoretical work on the merger of galaxy clusters suggested that 
the X-ray emitting gas will be strongly impacted during the interaction 
and that cooling flows may be disrupted by mergers though these 
calculations did not treat the fluid of the intracluster medium 
\citep{mcglynn84}.  While investigating the cluster Abell 2256, 
\citet{fabian91} concluded that this particular cluster is undergoing a 
merger currently and suggested that the accreted subcluster may donate
pre-cooled gas to seed a future, massive cooling flow.  In a series of 
controlled numerical experiments, \citet{burns97} and \citet{gomez02} 
examined the fate of an idealized, spherical cooling flow cluster in 
collision with another cluster for a variety of cluster mass ratios 
($1 / 4$ and $1 / 16$), cooling flow strengths (100 and 400 $M_{\odot} 
\mathrm{yr}^{-1}$) and central densities of the accreted subcluster 
($1.5 \times 10^{-3}$ to $1 \times 10^{-4} \: \mathrm{cm}^{-3}$).  They 
found that the ram pressure of the merging subcluster impacting the
cooling flow is the main mechanism for disrupting cooling flows.  However, 
the work of G\'{o}mez \textit{et al.} treated the hydrodynamics in two 
dimensions and considered only collisions at zero impact parameter 
(head on), thus simulating the most efficient case for
disruption.  Furthermore, their simulations treated the merging system 
in isolation from the larger, supercluster environment.  Given the 
simplifications made, these previous results do not reflect the rich 
complexity of interactions observed in both cosmological simulations and 
in observations.  In more recent work, fully three-dimensional cluster
collision calculations have been performed by \citet{ritchie02} and
\citet{ricker01}.  In both investigations the authors find that major 
mergers disrupt cooling flows.

The formation and evolution of cool cores in a realistic environment
can be investigated via numerical simulations that evolve both a 
collisionless dark matter component and a fluid that approximates baryonic 
matter from cosmological initial conditions to the present.  At a minimum, 
the numerical algorithm must incorporate energy loss from the fluid by radiation.  
The majority of numerical simulations that treat radiative cooling by the 
baryonic component have used the smoothed particle hydrodynamics (SPH) 
formalism.  Early SPH simulations with radiative cooling, starting with 
the work of \citet{thomas92} and \citet{katz93}, already found rough 
agreement with observed systems given the limited resolution possible at 
the time.  However, additional physical processes, supernova feedback in 
particular, were hinted at being necessary to avoid the ``cooling 
catastrophe'', whereby unrealistically large amounts of fluid would cool.  
The next generation of SPH calculations \citep{evrard94,frenk96,suginohara98} 
seemed to confirm the need for additional physics in the simulations and 
helped to quantify the strong resolution dependence of the over-cooling 
problem in SPH simulations.  In \citet{suginohara98}, for example, doubling 
the spatial resolution initiated substantial over-cooling that led to an 
improbably large X-ray luminosity for their cluster, given the cluster's 
average temperature.  Star formation was again found to be necessary to prevent
the divergence of the central density and X-ray luminosity in the recent,
high resolution simulations by \citet{valdarnini02}.
For the SPH simulations of \citet{pearce99,pearce00}, 
realistic clusters were found (with radiative cooling actually lowering 
the total X-ray luminosity for their clusters) provided that their SPH 
code was modified as follows.  First, cold and hot gas phases were 
identified and gas in the cold phase was excluded when calculating 
average density values in nearby regions of hot gas.  Second, a minimum mass 
limit for cooling was enforced and the resolution was set to an optimal 
value.  At higher resolution, however, over-cooling is expected to again 
contaminate their results.  Star formation was incorporated in one of the 
simulations of \citet{lewis00} where they examined the properties of a 
Virgo-like cluster at the present epoch and concluded that cooling with 
star formation eliminated the problems evidenced in \citet{suginohara98}.
However, in their simulation Lewis \textit{et al.} did find
a very large fraction of gas condensing (up to 30\%, roughly a factor of 
three too large compared to observations) despite the energy input from 
supernova feedback.  The phenomenon of over-cooling in SPH simulations has 
recently been cast in a very interesting light by \citet{tittley01}.  Briefly, 
a numerical ``drag'' was demonstrated as cold, condensed material traverses 
a hotter medium, causing spurious over-merging and over-cooling in SPH 
calculations.

In grid-based (Eulerian) treatments of the hydrodynamics of structure
formation, radiative cooling has been included in simulations calculated
with a first order accurate scheme \citep{cen92} and recently
with a more accurate total variation diminishing (TVD) scheme \citep{cen99}.
While it is much more straightforward to incorporate additional
physics such as star formation and multiple fluid components
in grid-based programs, these methods have historically suffered from
a lack of spatial resolution when implemented on static, uniform grids.
Compared to the resolution of current SPH simulations in dense regions, 
previous fixed grid Eulerian simulations have about an order of magnitude 
coarser spatial resolution (roughly 100 kpc in the best cases) if they 
are to evolve a useful subdomain in the universe.  For such simulations, 
the regions where radiative cooling dominates (roughly 100 kpc) would be 
unresolved with even a more accurate numerical scheme such as PPM.  The 
partial differential equations of hydrodynamics may, however, be solved 
on a sequence of nested grids of finer and finer resolution to allow a 
large range of length scales to be treated. The grid system must be dynamic 
to follow hierarchical collapse of the fluid during structure formation.  We 
use just such an adaptive mesh refinement (AMR) algorithm to attain the 
requisite spatial dynamic range to resolve cool cores while retaining the many 
desirable attributes of Eulerian hydrodynamics techniques.

In the current paper we address the issue of the formation and evolution of 
cool cores of gas in clusters through direct numerical simulations that 
proceed from realistic initial conditions.  We introduce  radiative cooling
by the fluid component into our algorithm and evolve clusters from cosmological
initial conditions through the non-linear hierarchical assembly phase to the 
present day.  We are thus able to observe merger processes in the simulated 
clusters where the full complexity of the system - ram pressure stripping, 
tidal disruption and shock heating - are treated accurately.

In \S 2 we describe our numerical code noting physical processes
that are not included in the current work and their potential importance.
In \S 3 we present our simulation sample, their gross physical properties 
and compare them to results from other simulations that include the effect 
of radiative cooling.
We then present a brief analysis of a test case where 
we have performed otherwise identical simulations with and without radiative 
cooling in \S 4.  In \S 5 we present an overview of subclusters in our simulations
and in \S 6 we present a history of cluster interactions
from a redshift of one to the present.  The impact of the cluster environment
and subcluster mergers are detailed in \S 7.
We then analyze our simulated data from 
an observational point of view in \S 8. 
Finally we compare our simulation results 
to recent observational results from \textit{Chandra}, and 
\textit{XMM} in \S 9 and summarize 
our conclusions in \S 10.

\section{Numerical Technique}

Our simulation tool is a coupled N-body Eulerian hydrodynamics code 
\citep{norman99,bryan01}.  We use an adaptive particle-mesh, N-body code 
to evolve dark matter particles.  To treat the fluid component, we use 
the PPM scheme \citep{colella84} on a comoving grid.  We employ adaptive 
mesh refinement (AMR) to concentrate our numerical resolution on the
collapsed structures that form naturally in cosmological simulations.  
The dark matter particles exist on the coarsest three grids; each subgrid
having twice the spatial resolution in each dimension and eight times the
mass resolution relative to its parent grid.  At the finest level, each
particle has a mass of $9 \times 10^{9} \; \mathrm{h}^{-1} \; \mathrm{M_\odot}$.
We use second order accurate TSC interpolation for the adaptive particle
mesh algorithm.  Up to seven levels of refinement are utilized for the fluid
component for most of the results presented here, yielding a peak resolution 
of $15.6 \: \mathrm{h}^{-1}$ kpc within our simulation box with sides of 
length 256 $\mathrm{h}^{-1}$ Mpc at the present epoch.  We have chosen 
a standard, flat $\Lambda$CDM cosmology with the following parameters:  
$\Omega_{\mathrm{b}} = 0.026$, $\Omega_{\mathrm{m}} = 0.3$, 
$\Omega_{\Lambda} = 0.7$, $\mathrm{h} = 0.7$, and $\sigma_{8} = 0.928$.

The fluid is allowed to radiatively cool in the present simulations.  We 
use a tabulated cooling curve \citep{westbury92} for a plasma of fixed, 
0.3 solar abundance to determine the energy loss to radiation.  The cooling 
curve falls rapidly for temperatures below $10^{5} \; \mathrm{K}$ and is 
truncated at a minimum temperature of $10^{4} \; \mathrm{K}$.  Heat transport 
by conduction is neglected in our algorithm and it is expected that even 
a weak, ordered magnetic field can reduce conduction by two to three orders of 
magnitude from the Spitzer value \citep{chandran98}.  Indeed, \citet{vikhlinin01} 
find that thermal conduction is suppressed by a factor of 30 to 100 at the 
interface of the ISM for NGC 4874 in the Coma cluster.  While Vikhlinin 
\textit{et al.} find that conduction is suppressed, the energy transport into 
NGC 4874 approximately balances the energy lost through radiation and therefore 
thermal conduction should be included in simulations that treat the detailed 
state of small scale structures.  Furthermore, the theoretical work of Narayan
\& Medvedev (2001) has shown that if turbulence extends over a sufficiently
large range of length scales within the ICM, the effective conductivity
coefficient is expected to be about one fifth of the Spitzer value making
thermal conductivity significant to the global energy balance of the ICM.
The potential importance of including conduction
in future work is also bolstered by recent data from
Fabian, Voigt \& Morris (2002) who report that in a sample of 29 clusters
most have an effective conductivity between the Spitzer value and one-tenth
the Spitzer value and from Zakamska \& Narayan (2003) who show that half of
the 10 clusters they studied are consistent with heat transport by conduction
balancing radiative cooling.

We also neglect energy input into the fluid from supernovae feedback or discrete
sources such as AGN in our current simulations.  \citet{lewis00} have found that
supernova feedback  and cooling can significantly alter the fluid properties in 
the centers of clusters though the large scale temperature distribution depends 
predominantly on shock heating and is relatively insensitive to both supernova 
feedback and radiative cooling \citep{dave01}.  
The expected effect from supernova feedback on our results is two-fold.
First, by neglecting additional
sources of energy such as supernova feedback our simulations 
generate cool cores with density profiles that are too steep and thus we will 
overestimate the resilience of cool cores during mergers.  
Second, it is precisely the cool material in cores that can collapse
to form assemblies of stars, removing that fluid from the flow and yielding
a less dense core.
Nonetheless, the 
case of maximal stability for cool cores is of theoretical interest as a
limiting case and also as a guide to constrain the largely uncertain 
parameterization of supernova feedback prescriptions.

As described in \citet{loken02} we have identified clusters within our 
$256^{3} \, \mathrm{h}^{-3} \, \mathrm{Mpc}^{3}$ parent volume and performed 
AMR simulations of sub-volumes to create a ``catalog'' of simulated clusters.  
From this set of adiabatic simulations we present two clusters, chosen because 
of their interesting merger history, that have been rerun with radiative 
cooling included.  One cluster, hereafter C1, undergoes a major merger between 
approximately equal mass components at $z \approx 0.4$ while our second simulation, 
hereafter referred to as C2, merges with a subcluster of approximately half 
its mass.  In future work, we will present results from an ensemble of radiative 
cooling clusters to address issues such as the cluster $\mathrm{L}_{\mathrm{X}} 
- \mathrm{T}$ relation and the probability of cool core disruption by cluster 
collisions.  For our present purpose we will limit our scope to these two 
clusters as a guide to interpreting the wealth of recent X-ray observations of 
galaxy clusters and to emphasize the importance of mergers in assembling cool
cores in rich galaxy clusters.

\section{Rich Cluster Properties}

\begin{figure}
\plotone{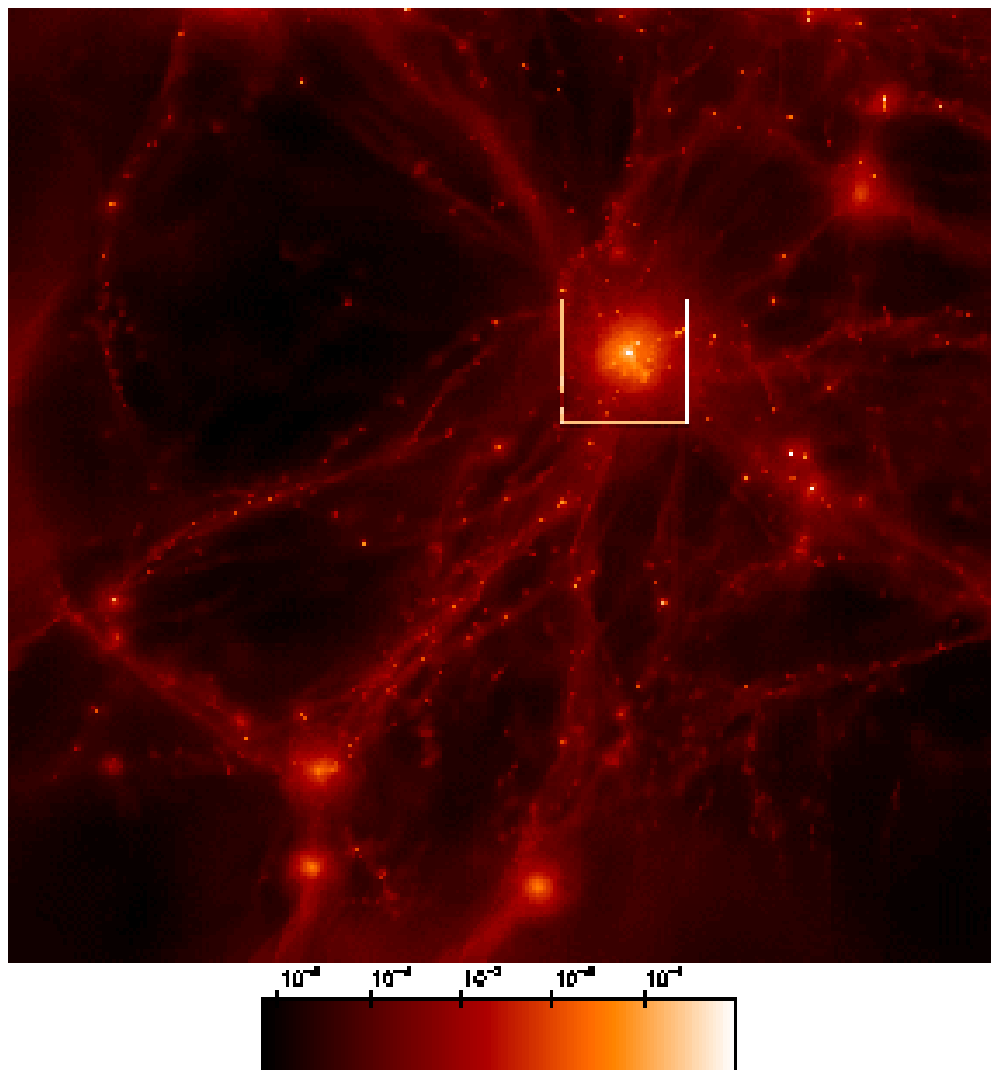}
\figcaption[f1.eps]{Projected gas density at the present day for the
  refined volume of C1.  The highlighted square marks a 5 $\mathrm{h}^{-1}$
  Mpc region around the dominant cluster. The entire image covers a region
  37 $\mathrm{h}^{-1}$ Mpc by 36 $\mathrm{h}^{-1}$ Mpc.  The color bar at
  bottom indicates the relative range in surface density. \label{fig:clrc00_ref_vol}}
\end{figure}

\begin{figure}
\plotone{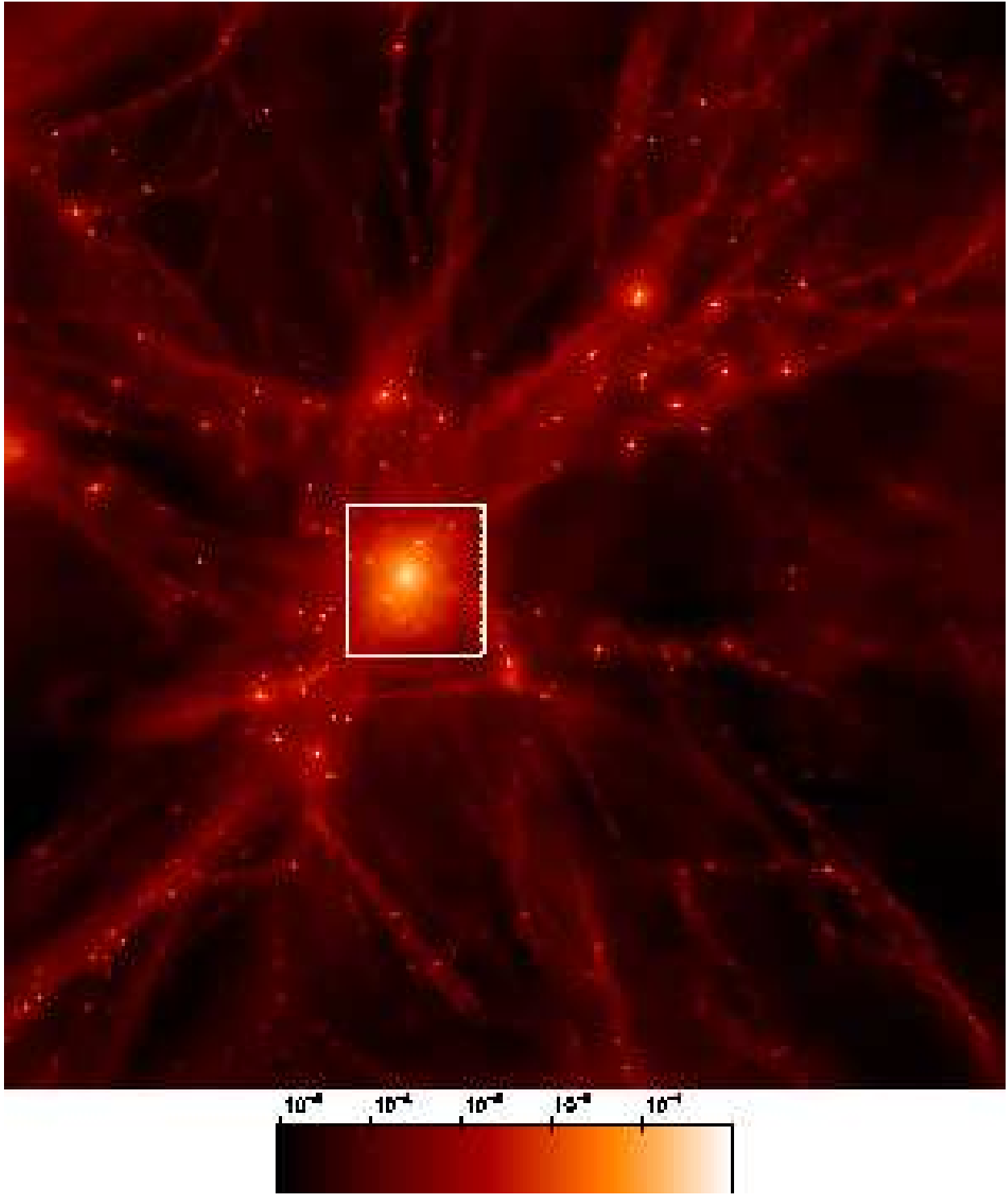}
\figcaption[f2.eps]{Projected gas density at present for the refined
  volume of C2.  The highlighted square marks a 5 $\mathrm{h}^{-1}$ Mpc
  region around the dominant cluster. The entire image frames a region
  35 $\mathrm{h}^{-1}$ Mpc by 38 $\mathrm{h}^{-1}$ Mpc.  The color bar
  at bottom indicates the relative range in surface density.
  \label{fig:clrc01_ref_vol}}
\end{figure}

Projections of the gas density at the present epoch within the refined volumes for
our simulated clusters C1 and C2 are shown in Figures \ref{fig:clrc00_ref_vol} and
\ref{fig:clrc01_ref_vol} respectively.  In both cases we see that the dominant 
cluster lies at the intersection of a network of filaments that convey material into 
the clusters.  In both figures we have outlined a square region 5 $\mathrm{h}^{-1}$ 
Mpc on a side about the center of mass of each cluster.  In subsequent images 
we will be zooming in on this region to examine the detailed structure present in 
our clusters.
The clusters we present here are the two most massive structures that form in our
simulation volume and are exceptional in that respect.  They correspond to
$R \ge 4$ and $R \ge 3$ clusters for C1 and C2, respectively, given the cluster
mass function in \citet{bahcall93}.  In adiabatic simulations of the clusters, they
exhibit a central cooling time shorter than the Hubble time (except for short intervals
around major mergers) during the simulations and would therefore be expected to
posses cool cores.

In Table \ref{tab:clusters}  we list gross parameters for the clusters C1 and 
C2 at a redshift 
of zero.  The virial radius, $R_{\mathrm{virial}}$, is calculated for an 
overdensity $\delta \rho / \rho$ of 200.  $M_{\mathrm{virial}}$, $M_{\mathrm{dm}}$, 
and $M_{\mathrm{gas}}$ are the total mass, total mass of dark matter particles, 
and total fluid mass within the virial radius, respectively.  Similarly, 
$L_{\mathrm{x}}$ is the total X-ray luminosity (line and continuum) within 
$R_{\mathrm{virial}}$ in the 1 to 10 keV band assuming a 0.3 solar metal abundance.  
The average physical temperature of the ICM within the virial radius is listed as
 $\mathrm{\overline{T}}$.
The velocity dispersion, $\sigma$, is the average, three-dimensional
velocity dispersion for the dark 
matter particles within the virial radius
and $f_{\mathrm{cool}}$ is the ratio of gas mass colder than 15,000 
K to the total gas mass within the virial radius.

The properties of our clusters are 
physically reasonable when compared to actual clusters.  In particular, unlike 
the high resolution simulation of \citet{suginohara98}, we do not find our
clusters to be overluminous for their temperature and do not conclude that additional
physics is required to offset radiative cooling based on gross cluster properties 
alone.  We note that our simulations have a peak resolution (grid cell size) less 
than half the size of the smoothing length for Suginohara and Ostriker's high 
resolution (MR) run.

We also find that the fraction of cool gas in our clusters, 8.5\% and 11.7\%, 
is in excellent agreement with the observational data for clusters summarized 
by \citet{balogh01} and is in reasonable agreement with the numerical results 
of \citet{pearce00}.  We have also rerun our C2 simulation with 6 and 8 levels
of refinement (corresponding to peak spatial resolutions of $31.2 \;
\mathrm{h}^{-1} \; \mathrm{kpc}$  and $7.8 \; \mathrm{h}^{-1} \; \mathrm{kpc}$
respectively).  As would be expected, the coarsest simulation produced a cluster
with a low fraction of cool gas, 2\%, this resolution not being sufficient to
accurately treat condensed cores.  For the highest resolution simulation, the
fraction of cool gas within the virial radius is 9.9\% - nearly identical to the 
result listed in Table \ref{tab:clusters} for our 7 level calculation.  
Other cluster properties 
are similar between the 7 and 8 level simulations.  We would like to 
emphasize that we do not see large $f_{\mathrm{cool}}$ values for rich clusters as has 
been reported in recent smoothed-particle hydrodynamics (SPH) simulations 
\citep{pearce00, lewis00, suginohara98} nor do we find that the fraction grows 
as our resolution improves beyond that used for the results presented here.

\section{Comparison to Adiabatic Simulations}

\begin{figure}
\plotone{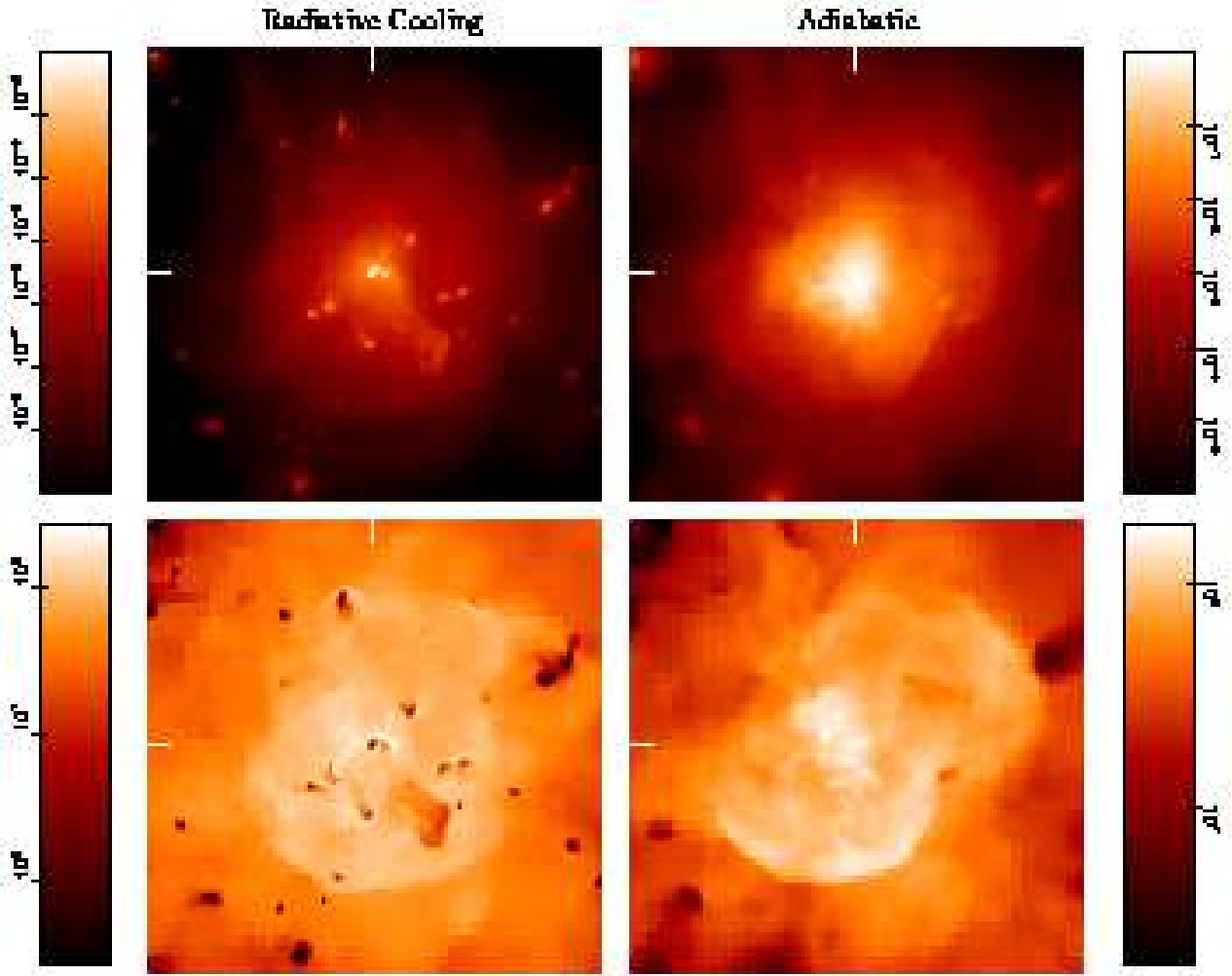}
\figcaption[f3.eps]{In the top row, we show the normalized, X-ray surface
  brightness image (in the 1 to 10 keV band) and at bottom the projected 
  temperature map for radiative cooling and adiabatic realizations of C1 
  at a redshift of zero.  The images show the central $5 \; \mathrm{h}^{-1} 
  \; \mathrm{Mpc}$ region.  The color bars indicate the range of surface
  brightness and emission-weighted temperature in Kelvin for each image.
  The tick marks above and to the left indicate the cluster center of  mass.
  \label{fig:cl00_rc_adiabatic}}
\end{figure}

We have, as a test, performed simulations of the same refined volumes in
the adiabatic limit and with radiative cooling.   In Figure 
\ref{fig:cl00_rc_adiabatic} we show projections of the X-ray luminosity (top row) 
and emission-weighted temperature (bottom row) for the adiabatic and radiative 
cooling realizations of C1 at the present epoch.  The image scales are
indicated by the color bars and 
and the surface brightness has been normalized
to its maximum value.  The most noticeable distinction is 
that when radiative cooling is included, subclusters form dense cores of cool 
gas (bright and dark blobs in the X-ray and temperature images, respectively) 
as expected and seen in previous simulations \citep[c.f.][]{katz93}.  In 
particular note the tight, high density core of cool gas at the center of the 
primary cluster (the cluster center of mass is indicated by the tick marks
for reference).  
As infalling cores interact with the primary cluster, they produce complex 
structures such as the bow shock and ``comet tail'' associated with the subcluster 
to the bottom of the cluster center and the extended, irregular distribution of 
cool gas (``cool front'') to the bottom right of the cluster center.  Without 
the density enhancement generated by cooling, the collisions of clusters in the 
adiabatic simulations lack this rich structure.

The larger scale structure of the adiabatic and cooling clusters are generally 
similar, however, as they are dictated by the overall cluster properties rather 
than perturbative interactions.  For example, the ridge of shocked, hot gas at 
the bottom of the clusters is in nearly the same location in both the radiative 
cooling and adiabatic simulations and has comparable temperatures in both cases.

\begin{figure}
\plotone{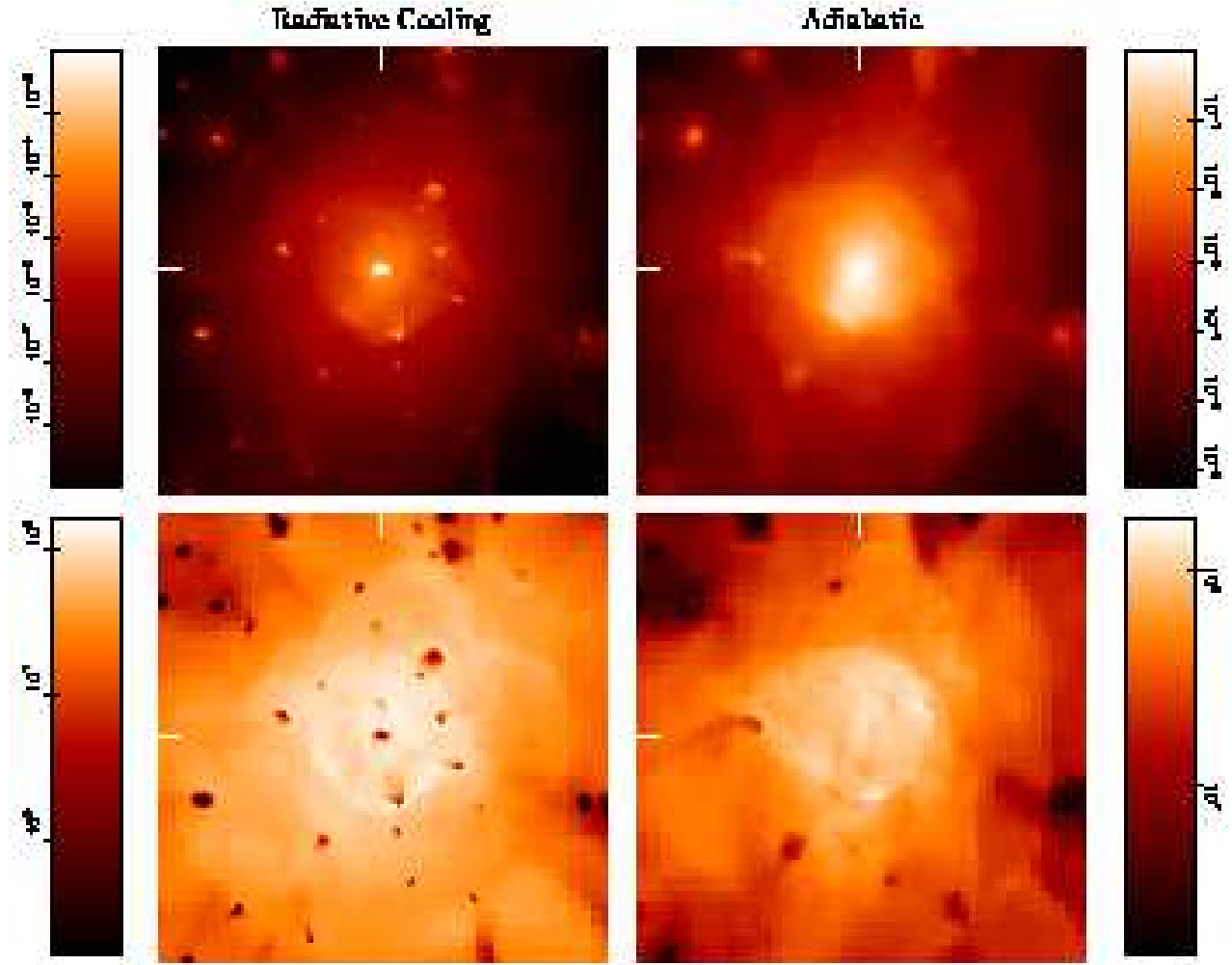}
\figcaption[f4.eps]{Projected, normalized X-ray surface brightness (top row)
  and projected temperature maps for C2 at a z = 0.25.  The images show the 
  central $5 \; \mathrm{h}^{-1} \; \mathrm{Mpc}$.  The center of mass of the 
  central cluster is indicated by the tick marks.  The color bars are as in 
  Figure \ref{fig:cl00_rc_adiabatic} \label{fig:cl01_rc_adiabatic}}
\end{figure}

As another example, Figure \ref{fig:cl01_rc_adiabatic} shows images 
for the C2 simulation at a redshift 
of 0.25.  Again, a core of dense cool gas is readily apparent in the radiative 
cooling results though the core is more regular in this instance.  This 
snapshot also demonstrates cluster interactions; a subcluster can be seen 
below the cluster center of mass with a leading shock front and a trailing 
stream of stripped, cooler gas.  While the filamentary network is bringing 
material into the cluster in the same manner in both the radiative cooling 
and adiabatic simulations, the infalling subclusters are much more condensed 
in the radiative cooling simulation and are hence more readily apparent.
The more strongly peaked density profile
also serves to shield the infalling cluster from disruption.  This core 
of cool gas has survived a cluster crossing  with only some ram pressure 
stripping and tidal stripping in evidence.  In the adiabatic case, this 
infalling cluster was absorbed into the primary ICM with only a relatively 
small elongation of the central surface brightness distribution and small shock front 
at the bottom of the cluster to betray its presence.

\begin{figure}
\plotone{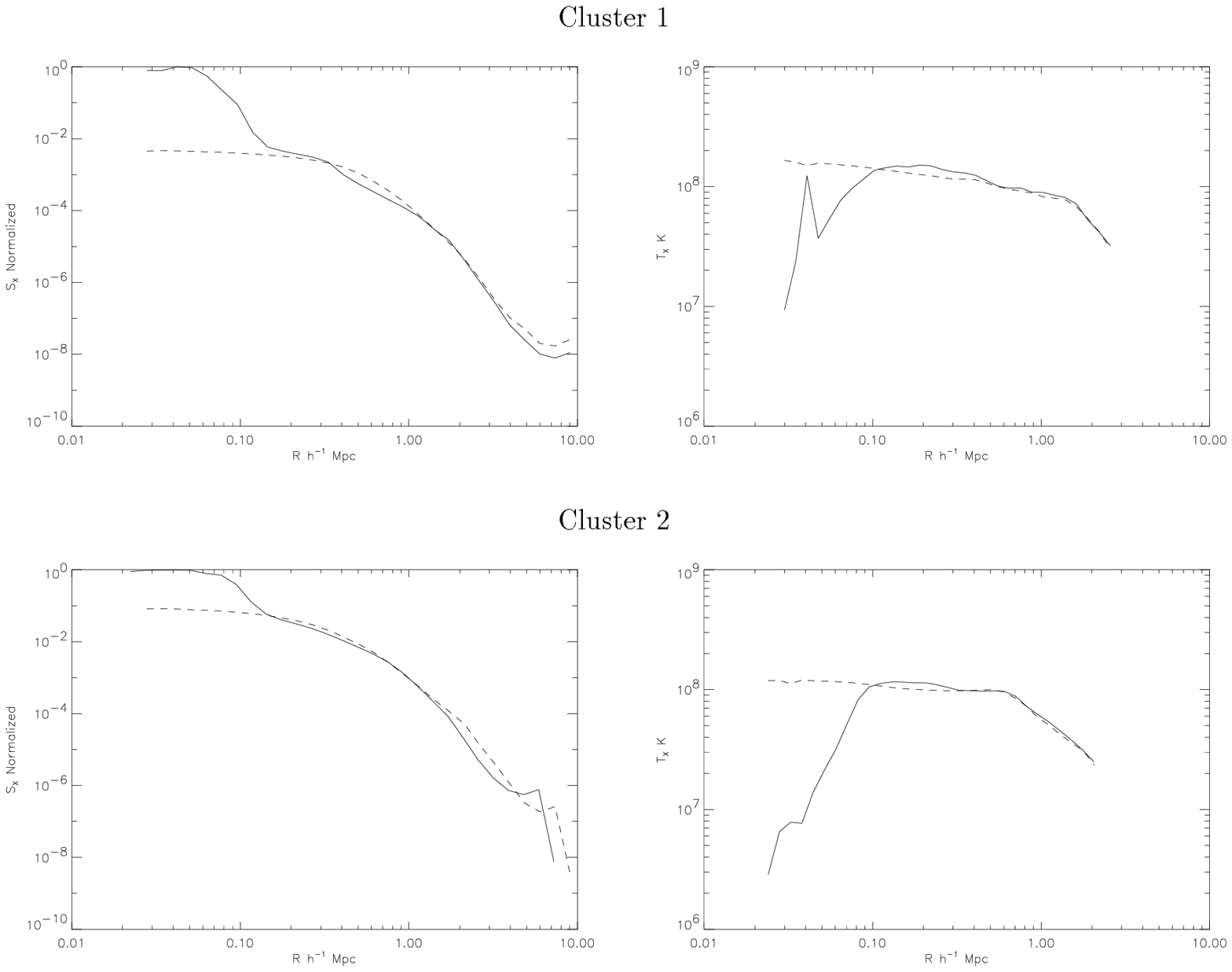}
\figcaption[f5.eps]{Profiles of the normalized X-ray surface brightness and
  projected luminosity-weighted temperature for C1 at $z = 0$ (top row) and
  C2 at $z = 0.25$ (bottom row).  The solid curves correspond to the simulation
  with radiative cooling and the dashed curves are from the adiabatic simulation.
  \label{fig:profile_rc_adiabatic}}
\end{figure}

In Figure \ref{fig:profile_rc_adiabatic} we show profiles of the X-ray surface 
brightness and the projected, X-ray emission-weighted gas temperature 
corresponding to the snapshots shown in Figures \ref{fig:cl00_rc_adiabatic} and 
\ref{fig:cl01_rc_adiabatic}.  The results from the adiabatic simulations are 
drawn as dashed curves while the radiative cooling profiles are solid lines.
Both C1 and C2 show an excess X-ray luminosity and falling temperature within 
about 100 ${\mathrm{h}}^{-1}$ kpc of the cluster center, as would be expected for 
a cool core.  Outside of the core there is little difference between the adiabatic 
and radiative cooling simulations in spite of the rich structure of cool gas clumps
in the images.  Also note that the gas in the core is not very cool, 
remaining above a million Kelvin for both simulations.  As noted in \S 2 our cooling 
function continues down to approximately $10^{4}$ K meaning that the fluid can lose 
energy by radiation until this floor value is reached (neglecting temperature changes 
due to mechanical work on the fluid).  The high temperatures of the ``cool'' gas in 
our cooling clusters is due to two effects.  First, we have shown the profiles 
derived from a projection of the temperature field onto the plane of the sky.
The cool core is seen along with the hotter material lying along the line of
sight.  Second, as will be discussed in more detail later,  the physical temperature 
of the core reaches values above $10^{6}$ K.  There are only two means of heating
possible given the assumptions we have made for these simulations: shock heating and 
adiabatic compression.  We find this to argue for the importance of accurately 
treating shocks in the ICM.

\section{Subcluster Properties}

The vast majority of cluster interactions in our simulations can be described as
accretion of smaller subclusters by the dominant cluster.  In this section we
detail the properties of these subclusters.  In Table 2 
we list gross 
properties for a sample of outlying subclusters that reside between 2 and 5 $\mathrm{h}^{-1}$
Mpc of C1's center of mass at a redshift of zero.  Of these 13 subclusters,
ten are falling in for the first time.  Two of the subclusters (the first
and third entry in Table 2) are climbing out of the cluster potential
well at the current epoch 
and one subcluster (the fifth entry) has orbited
about the main cluster for the past six billion years at a typical separation
between 2 and 3 Mpc.

In Table 2, we list the distance from the C1 center of mass
which helps gauge how well the subclusters
may be treated as isolated systems.  We also list the virial
radius, virial mass, the total dark matter and fluid mass and X-ray
luminosity within $\mathrm{R_{virial}}$.  We tabulate the average
physical temperature of the gas within the virial radius ($\overline{\mathrm{T}}$)
and within a core of 50 kpc ($\mathrm{T_{core}}$), the average, 
three-dimensional velocity dispersion of dark matter within the
virial radius, the fraction of gas that is cooler than 15,000 K
within the subcluster, the average cooling time within the core and
the virial ratio for the subcluster.  The virial ratio is computed
as $\sigma^{2} \mathrm{M_{virial}} / \left| W \right|$ where $\sigma$
is the previously described velocity dispersion and $W$
is the gravitational potential energy calculated from spherically
averaged profiles of the dark matter distribution from
\begin{equation}
   W = - 4 \pi G \int_{0}^{\mathrm{R_{virial}}} \rho \left( r \right)
      M \left( r \right) r dr.
\end{equation}
The virial ratio is expected to be approximately unity for
isolated structures in virial equilibrium.  
The two subclusters with a measured virial ratio greater than 2
have already interacted with the main cluster before this time
and in particular, the fifth subcluster (with a virial ratio
of 4.1) has orbited within the clusters virial radius for a
significant fraction of the simulation.  The pristine, infalling
subclusters are all relatively well described as gravitationally bound systems
in virial equilibrium.

Overall, the subclusters have properties similar to galaxy groups
(in terms of extent, mass, average temperature and luminosity).
Their baryons are locked in cool, condensed gas; all subclusters
have low core temperatures and high fractions of cool material.
Furthermore, the cooling time is much shorter than the Hubble
time for all subclusters in Table 2.  In our simulations we have
only included the effect of radiative cooling without any heat
sources or transport  mechanisms beyond the adiabatic physics
level.  For this set of assumptions, an isolated subcluster
has no recourse but to cool a large fraction of its baryons.

Given the large fraction of cool material, star formation feedback
will obviously impact the detailed structure of these subsystems.  
They will, however, likely remain as 
gravitationally bound structures and act as sources of condensed
gas for the rich clusters that form from them.
As mentioned previously, we
are focusing our attention on the properties of rich clusters
and considering only the limiting case where radiative cooling is
as efficient as possible.

\section{Merger Histories}

\begin{figure}
\plotone{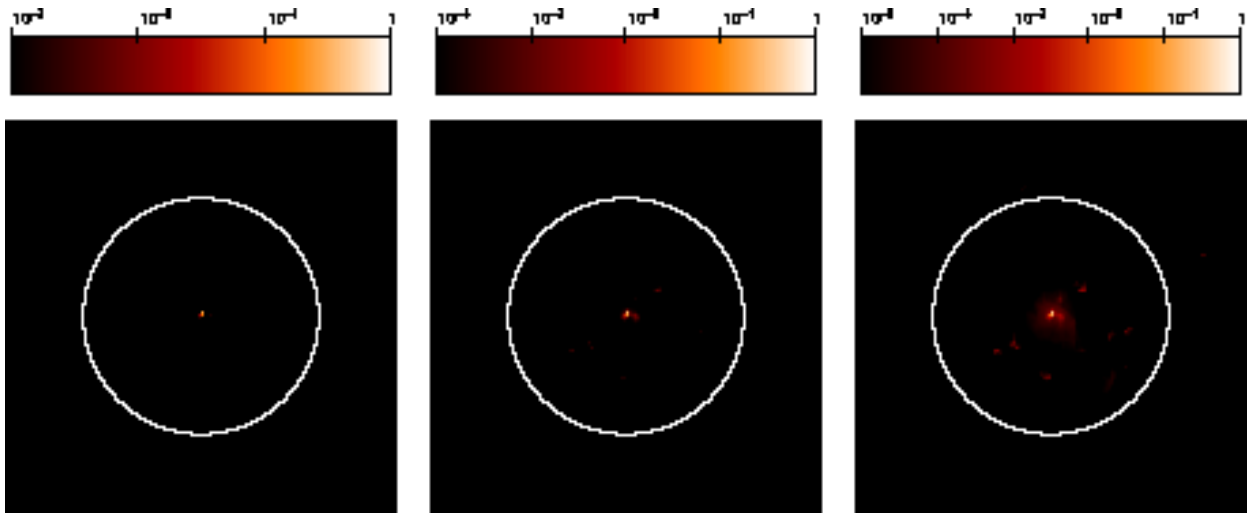}
\figcaption[f6.eps]{The normalized X-ray surface brightness for C1 at a 
  redshift of zero for three image scales (from left to right, the minimum 
  is $10^{-3}$, $10^{-4}$, and $10^{-5}$ of the peak intensity).  A circle 
  of radius $1.5 \; \mathrm{h}^{-1} \; \mathrm{Mpc}$ and centered on the 
  cluster center of mass is overlayed on the data.  The X-ray emission is 
  very strongly peaked at the cluster center, with a dynamic range of $10^{4}$ 
  only one subcluster is visible in the image and several other subclusters 
  are visible if the dynamic range is extended to $10^{5}$.
  \label{fig:xray_range}}
\end{figure}

In this section we present a detailed discussion of the history of our simulated
clusters.  In Figure
\ref{fig:cl00_table} we show projected, normalized X-ray surface brightness
and X-ray emission-weighted
temperature maps for C1 from a redshift of one to the present day in approximately
500 million year intervals.  The images have all been scaled to the same range of
values so that a given color corresponds to the same value (within the discreteness
of the color map) throughout the sequence of images.  
We note that the X-ray images have been prepared to emphasize the rich substructure
by firstly showing all pixels with a surface brightness 1\% or greater than
the peak as white.   Secondly, the dynamic  range extends
far beyond that possible with current observations
(please note the color bars in Figure \ref{fig:cl00_table}).  
On this scale, there is a very rich array of subclusters and substructure
visible.  In Figure \ref{fig:xray_range} we show the X-ray surface brightness
for C1 at the present epoch with three example image scales.
For a more realistic dynamic range of $10^{3}$ there is only one structure feature
visible within 1.5 Mpc of the cluster center of mass.
For observations with a dynamic range of $10^{4}$ there
are two subclusters visible within an Abell radius while the remainder of 
projected subclusters and substructure seen in Figure \ref{fig:cl00_table} 
are fainter still.

At $z = 1$ the primary has
a regular core of cool gas though it is interacting with a set of irregular
subclusters to the lower left and a comparable mass cluster is assembling to the
far right (in projection).  Through a redshift of 0.75 the primary core is bar shaped
and there are several filaments of cool, dense gas running through the cluster.

Interactions continue through a redshift of 0.5 when the central cool core 
is noticeably reduced in extent in the temperature map.  At this epoch the cluster
to the right is still taking shape and is moving closer to the primary through
$z = 0.43$ and finally collides nearly head on with the primary at a redshift of
0.37.  There is a collar of shock-heated gas surrounding the core at this epoch
which expands outwards.  The impactor reaches its turning point after passing
through the cluster at approximately $z = 0.3$.  There is a strong shock front 
at the head of the impactor and a trail of cooler gas leading back from the
shock front to the cluster center.  Surprisingly, a core of cool gas is still
visible at the cluster center of mass.

By $z = 0.25$ the impactor has fallen back towards the center and there is a binary
cool core system that persists through $z = 0.21$.  At $z = 0.25$ the shock
fronts continue to expand away from the cluster center.  The binary cores have
merged by $z = 0.16$ and the core is rather small in extent.  From $z = 0.12$ to 0.08
a smaller subcluster passes through from left to right across the cluster and
from $z = 0.08$ to the present a set of subclusters fall through the cluster from
the upper left to the bottom right, one of which produces a significant bow
shock ahead of it and is finally disrupted at $z = 0.0$ into an irregular patch
of cool gas (also see Figure \ref{fig:cl00_rc_adiabatic}
for an enlarged view of this cluster at $z = 0$). 

Similar images for C2 are shown in Figure \ref{fig:cl01_table}.  The C2 sequence
begins as the primary cluster has merged with a subcluster of about half its
mass. The cores of the two systems nearly overlap one another at the cluster center.
There are, in addition, several smaller scale interactions lasting to $z = 0.65$.
From $z = 0.65$ through $z = 0.43$ the primary cluster is left largely unperturbed.  Nonetheless,
the central cool core is almost totally absent at $z = 0.5$.  From $z = 0.37$ through to
$z = 0.12$ a cluster with about 5\% of the mass of the primary falls from top to
bottom through the cluster.  It passes through the cluster center between redshifts of
0.31 and 0.25 and excites a spherical shock wave that expands through the cluster from 
redshifts of 0.25 to 0.16.  The infalling cluster is disrupted by the encounter, ultimately
becoming the ``Y'' shaped structure below the primary cluster in the snapshots at 0.16 and
0.12.  Excepting a few minor, off axis,  interactions the C2 system is left largely undisturbed from
z = 0.25 to the present.  It is interesting to note that the central cool core does not
regain a spherical shape in this intervening time, remaining instead bar-like in this
projection.

As a means of gauging the disruption of the cool cores we plot the X-ray surface
brightness and spherically averaged temperature profiles for clusters C1 and C2 at
a variety of redshifts in Figures \ref{fig:cl00_prof_table} and \ref{fig:cl01_prof_table}
respectively.  The temperature profiles for both clusters are very complex,
though in all cases there is cool gas in a core of some form.  The temperature
curve for C1 at $z = 0.25$ has an odd shape because there are two cool cores
in a binary at this epoch and the profiles are taken about the center of mass
of the cluster as a whole which lies between the two cores.  The smoothing over
the binary core system is also responsible for the isothermal profile shape for
the X-ray surface brightness of C1 at $z = 0.25$.  At $z = 0.31$, however, the X-ray
surface brightness signature of a cool
core is washed out by the recent passing of the companion cluster
through the primary.  At redshifts of $z = 1$ and 0.84, C1 again does not
exhibit an X-ray excess.  For C2 the surface brightness profiles are much more
straightforward; for all redshifts from 0.43 to 1 (with the exception of 0.57)
there is no evidence for an X-ray excess.  From 0.37 up to the present, C2 does
not suffer any major interactions and shows a remarkably static luminosity
excess from its cool core.

From an observational point of view, one may see clear evidence for
a cool core at only certain points in the cluster's history.  If there has
been a major merger event or frequent, though less extreme interactions,
the cool core signature may disappear in the X-ray surface brightness.   
However, cool gas remains in these two systems at all times, despite
the vigorous merger activity.

\section{The Dynamical Cluster Environment}

In this section we further explore the complex dynamical environment that
acts on the cool core clusters.  We note in particular that our simulation
results are inconsistent with expectations arising from the simple cooling
flow model for an isolated cluster.  The dynamical cluster environment is 
seen very clearly from the velocity field which we plot for 4 epochs for
cluster C1 in Figure \ref{fig:cl00_vel}.  These are representative examples
of the flow field; similar features are seen for both clusters and for
other slices through the clusters.  We plot the velocity in the
x-y plane for a $1 \; \mathrm{h}^{-1} \; \mathrm{Mpc}$ region centered
on the cluster center of mass.  At a redshift of 0.25, two cores are
about to collide with one another and a similar collision is seen at
a redshift of 0 where a subcluster has passed through the core.  In
all plots, the velocity field is irregular with material streaming
through the cluster core.  Recall that the clusters lie at the intersection
of filaments (see for example Figures \ref{fig:clrc00_ref_vol} and 
\ref{fig:clrc01_ref_vol}) that flow into the clusters.
The asymmetric streams result in a complex flow field that dominates over 
the orderly, spherically symmetric radial flow of material arising from
a ``cooling flow''.  In addition to the features exemplified in Figure
\ref{fig:cl00_vel}, we also see circulation within the cluster
center.  The rotation arises from the residual angular momentum of merging
structures colliding at non-zero impact parameter.

The response of the cool core is characterized in Figure 
\ref{fig:cool_parameters} where we plot the following quantities versus 
the lookback time (and redshift) for the C1 and C2 clusters: (1) the amount of cool gas 
(gas with temperature less than 15,000 K) within 100 $\mathrm{h}^{-1}$ kpc 
of the cluster center of mass, (2) the cooling radius, that is the radius where the 
cooling time equals the Hubble time at that particular redshift, (3) the 
characteristic core temperatures of the gas which we take to be the average
temperature within 50 kpc of the cluster center of mass, and (4) the fraction
of gas cooler than 15,000 K within the virial radius.  The characteristic
temperature is calculated from both a spherical average of the gas temperature
(solid curve) and from a circular region of 50 kpc radius
from the projected, emission-weighted temperature (dashed curve).
Recall from the discussion in \S 5
that C1 experiences a major merger that lasts from $z = 0.37$ to $z = 0.21$
and several minor interactions.  C2 experiences a major interaction 
at $z = 1$.

The mass of cool gas grows for both clusters, though it often increases by discrete 
units causing peaks in the plots of cool mass.  
For C1 at $z = 0.25$, the sharp drop in cool gas mass is
due to a binary core system orbiting the cluster center of mass beyond our
chosen aperture of 100 $\mathrm{h}^{-1}$ kpc for this measurement.
The mergers are complicated
phenomena, infalling cores both donate cool material to the core and
shock heat the ICM which mixes hot fluid into the core material.

The cooling radius shown in the  second row of plots in Figure 
\ref{fig:cool_parameters} also changes abruptly.  As dense, condensed
material accumulates in the core the cooling  time drops causing
the cooling radius to grow.  During the quiescent period for C2
(from Figure \ref{fig:cl01_prof_table}, the epochs where there is a static luminosity
excess, $z = 0.5$ to 0) the cooling radius grows slowly with time.
The characteristic core temperature is shown in the third row
of plots in Figure \ref{fig:cool_parameters}.  The projected temperature
is at nearly all times systematically higher than the physical core
temperature and responds less strongly
to merger events.  The characteristic physical temperature of the core
exhibits abrupt changes as shocks arising from the cluster collisions
heat fluid in and around the core.  We also note that the characteristic
projected temperature for the cores of C1 and C2 agree with recent
observations of cool core clusters where minimum temperatures have
been consistently found to be approximately one keV for a variety of clusters
\citep{fabian02}.  For a global view of the evolution of clusters C1 and
C2, we also show the fraction of gas cooler than 15,000 K within the
virial radius ($f_{\mathrm{cool}}$ from Tables 1 and 2).
Cool gas is present in both clusters throughout their evolution despite
collisions.  As
subclusters fall in (objects with a majority of their baryons in a cool
phase as shown in Table 2) most of their baryons are heated by shocks
and compression so that as the cluster grows in mass, the cool fraction
generally declines with time.  However, as noted previously, some cool
gas survives these interactions and merges in the cluster core as well.
As an example of the interaction process, a typical infalling cluster
at a redshift of 0.5 will lose $\approx 90\%$ of its total gas and 
typically $\approx 70\%$ of its core mass (the cool gas) has been
heated and lost to the cluster during a crossing (see Figures 7 and 8).
These results, of course, can vary widely depending on the trajectory
of the infalling subcluster and the state of both systems.

The main point of this examination of the cores is that the mass of 
cool gas in the central core grows predominantly from the accretion of discrete 
subclusters that contain pre-cooled gas and not from steady state cooling.  We
do not find any evidence for a ``cooling flow'' in our simulations, in accord 
with the recent simulations of \citet{lewis00} who find the flux of mass in 
adiabatic and radiative cooling simulations agree with one another in to a 
distance of 40 kpc from the cluster center.  For our C1 and C2 clusters, the 
random flow of fluid in the cluster center (characterized by the average 
velocity dispersion for the fluid within the cooling radius) exceeds
the mean radial velocity by a factor of approximately 25 (see 
also Figure \ref{fig:cl00_vel}).  
This is clearly inconsistent with the classical cooling flow scenario.

On average, C1 and C2 gain cool material at a rate of about 600 $M_{\odot} 
\mathrm{yr}^{-1}$ from a redshift of 2 to the present but this is predominantly 
due to merger events.  In Figures \ref{fig:clrc00_cold_flux} and 
\ref{fig:clrc01_cold_flux} we show the flux of cool material through spheres 
of radius 0.1, 0.5 and 1.0 Mpc centered on the cluster center of mass for C1 
and C2, respectively.  If, for example, a blob of material crossed through 
the cluster without interacting at a constant velocity of 1,000 
$\mathrm{km} \; \mathrm{s}^{-1}$ 
it would appear as a negative and then positive peak in these plots, separated 
in time by approximately 2 billion years.  One might also expect that as cool
cores rain in on the cluster they would appear in these plots as a set of three 
peaks as they successively cross from 1 to 0.5 to 0.1 Mpc.  Figures 
\ref{fig:clrc00_cold_flux} and \ref{fig:clrc01_cold_flux} show that the 
situation is more complicated than this expectation.  In Figure 
\ref{fig:clrc00_cold_flux} for example, there is little flux of cool
material through the surface at a 500 kpc while we can clearly see cores 
crossing at the other two radii.  
The situation for C2 is largely similar though we can see that 
this cluster has a more gentle evolution (note the different scales for 
Figures \ref{fig:clrc00_cold_flux} and \ref{fig:clrc01_cold_flux}).  From a 
redshift of approximately 0.6  to the present, C2 is left largely undisturbed 
and there is an indication that cool material is settling on the core given 
the small, approximately constant flux of material through the inner surface 
for the past 2 billion years.

As has been demonstrated several times, the evolution of these simulated 
cool core clusters  is complex.  Even at the present epoch, the clusters 
do not represent relaxed, steady state structures as can be seen from Figures
\ref{fig:cl00_table} and \ref{fig:cl01_table} for instance.  
The full complexity of the merger history of the
clusters is intrinsically tied to the formation of cool cores in our simulations.
This stands in stark contrast to the classical cooling flow scenario where a spherically
symmetric distribution of gas cools in isolation and, in particular, is not perturbed
by accretion or major merger events with other cool cores.

\section{Observational Analysis}

If we analyze our simulated clusters from the point of view of an observer, 
they would be identified as a ``cooling flow'' cluster if they exhibit an 
X-ray excess over an isothermal sphere model and/or have lower temperatures 
at the center.  No actual flow has ever been detected, nor could be
detected with current X-ray spectrometers.

In the simplest application, one ascribes all the excess luminosity to a cooling flow
according to the prescription \citep{fabian-araa-94}
\begin{equation}
   L_{\mathrm{cool}} = \frac{5}{2} \frac{\dot{M}}{\mu m} k_{B} T.
\end{equation}
We have fit isothermal sphere models to our surface brightness profiles
to determine the luminosity excess (surface brightness and best fit isothermal
models are shown at selected epochs for C2 in Figure \ref{fig:c2_mdot_excess}). 
Applying equation (2) to C2 (given the
known temperature distribution within the cluster) results in the mass deposition
rates shown in Figure \ref{fig:c2_mdot_excess}.  This method
gives reasonable mass deposition rates (in comparison to observations)
though as has been previously demonstrated, there is not
a real flow present in our simulations.

The hierarchical merger scenario for cool cores provides a physical basis for the results
of \citet{loken99}.  In their analysis they consider a nearly complete, volume-limited sample of nearby
($z < 0.1$) Abell clusters. To a high degree of significance, Loken \textit{et al.} find that massive cooling
flow clusters ($\dot{M} > 100 M_{\odot} \, yr^{-1}$) have closer nearest neighbors
and reside in more crowded environments than clusters without cooling flows.  If cool cores
in rich clusters
are built primarily through merger events, massive cool cores will have to
reside in dense environments for there to be a sufficient merger rate.  As seen in our
simulations, mergers can disrupt the X-ray signature of a cool core, though inevitably
the core is renewed.  Clusters with significant substructure may or may not show evidence
for a cool core, depending on the details of the particular merger history but cool
core clusters will be predisposed to reside in overly dense regions of the universe.

\section{Comparison to Recent Cluster X-ray Observations}

Recent, high resolution observations with \textit{Chandra} and \textit{XMM}
have revealed new levels of detail in galaxy clusters.  What was seen
with previous instruments to be relaxed, smooth clusters now show
evidence for a dynamic nature.  From Figures \ref{fig:cl00_table} and 
\ref{fig:cl01_table} we see that our
simulations also demonstrate a rich array of cluster substructure and
remnant signatures of interactions in the projected temperature
maps.

Many clusters have been found to contain ``cool fronts'' - clouds of cool
gas embedded in the hot ICM.  Examples include
Abell 2142 \citep{markevitch_2142}, Abell 3667 \citep{vikhlinin_3667}, RXJ1720+2638
\citep{mazzotta01}, Abell 85 \citep{kempner02}, the Centaurus cluster \citep{sanders02},
MS 1455.0+2232 \citep{mazzotta02},
Abell 754, and Abell 2163 \citep{markevitch_2163} which 
all show evidence of a contact discontinuity between an irregular distribution of 
cool gas and the ICM.  An example of a cool front 
in a cluster is shown in the  upper left panel of Figure  \ref{fig:obs_comparisson} 
where we reproduce data for Abell 2142 \citep{markevitch_2142}.
The temperature maps from figures
\ref{fig:cl00_table} and \ref{fig:cl01_table} provide evidence for a natural explanation
for the clouds observed in cool front clusters in terms of hydrodynamical effects
from cluster collisions.

A striking example of the complex structure of cool gas was recently 
discovered by \citet{fabian01}
in Abell 1795.  We show the adaptively smoothed \textit{Chandra} image of the core of 
Abell 1795,
a well known cool core cluster, in the upper right panel of Figure 
\ref{fig:obs_comparisson}.  
The image is approximately 80 kpc across.  From our simulations, we find that similar
filamentary structures can be formed from the relative motion of a condensed
core of cool material through a hot medium.

In our hierarchical merger model,  cluster  mergers  bring cool material
into the accreting cluster.  An infalling core was recently reported from
\textit{Chandra} observations of 1E0657-56 \citep{markevitch02} which is shown
in the bottom left panel of Figure \ref{fig:obs_comparisson}.  
They term the infalling system a ``bullet subcluster'', the bullet being a core
of cool gas (between two and three times cooler than the cluster average)
traversing the primary cluster with a prominent bow shock plowing through the 
cluster medium reminiscent of interactions seen frequently in our simulations.

At higher redshifts, cluster  interactions  become more frequent.  Our
cluster C2 is relatively quiescent from a redshift of a half to the
present - the bulk of its accretion occurring further in the past.
A relatively distant cluster is shown in the bottom right panel of Figure 
\ref{fig:obs_comparisson}.
At a redshift of $z = 0.407$, CL 0939+4713 (which was observed with \textit{XMM}
by De Filippis \textit{et. al} 2001)
shows significant substructure.
In the image there are several peaks in the central X-ray distribution
(though the peaks labeled P1 and P2 represent a foreground source and a background
quasar and do not  indicate substructure).  Their 1 arcminute bar corresponds to 
approximately 300 kpc for our chosen cosmological parameters.

\section{Conclusions}

From the simulation results presented here, we see that the formation and evolution of
cool cores in rich galaxy clusters is a highly dynamic process, even to the present
epoch.  The cluster appearance is shaped by
hydrodynamical interactions between condensed cores of cool gas and the hot intracluster
medium.  Because of these complex interactions, the central cool cores depart 
significantly from spherical symmetry and do not correspond to the picture evoked by
the steady state cooling flow model.

We find that accreted cool cores can survive a transit through the cluster and, more
importantly, can donate pre-cooled gas to the accretor.  Thus, as rich clusters are assembled
through the hierarchical merger scenario of the cold dark matter model, cool cores are
built simultaneously from the baryonic fluid component.  We note that our simulations have
only treated the case of maximal cooling in the sense that heat transport by conduction and
energy input are neglected.  Our accreted subclusters thus have a steeper density profile
than is expected for real clusters and this in turn overestimates the stability of merger
subunits in our simulations.  In future work we will investigate the impact of star formation
feedback and thermal conduction on our current simulation results.  
We have also only presented results
from the most massive clusters in our simulation archive.  In future work we will present a
detailed analysis of cluster-cluster and cluster-subcluster interactions seen in our complete
sample of simulated galaxy clusters evolved in the limit of radiative cooling only.

In both radiative cooling clusters presented in the current work, 
we see no evidence for a cooling flow despite the fact that we have sufficient
numerical resolution to marginally resolve a vigorous cooling flow if one were present.  We
instead find that cooling cores gain cool gas through discrete accretion events and that the
cluster's intrinsic velocity field would overwhelm the expected field implied by a 
cooling flow model.  We emphasize that the cool cores in our
simulations are dynamic systems, tightly coupled to events in the surrounding intracluster medium.

The model of hierarchical assembly of cool cores naturally accounts for the rich array
of structure reported in recent X-ray observations in terms of merger processes.  Likewise,
the model accounts for the prevalence of both cluster substructure
(evidence of merger activity) and cool cores in clusters of galaxies.  It also provides
a physical basis for the result that massive cool  cores are found preferentially
in dense supercluster environments \citep{loken99}.  
%We find that we can produce realistic, rich cooling core clusters with only
%radiative cooling included in our simulations.  In particular, our two target
%clusters have X-ray luminosities consistent with their mean temperatures and
%have characteristic (projected) core temperatures that agree with recent observations.

\section{Acknowledgments}

PMM wishes to thank Romeel Dav\'{e} for helpful comments on the manuscript.
The simulations presented in this paper  were carried out on the Origin2000 system
at the National Center for Supercomputing Applications at the University of  Illinois,
Urbana-Champaign with computer allocation grant AST010014N.

\onecolumn

\begin{deluxetable}{l || c | c}
      \tablewidth{0pt}
      \tablenum{1}
      \tablecolumns{3}
      \tablecaption{Cluster Properties \label{tab:clusters}}
      \startdata
      \hline
                            & C1
                            &  C2                                \\
      \hline
      \hline
      $\mathrm{R_{virial}}$
      & $2.7 \: \mathrm{Mpc}$
      & $2.4 \: \mathrm{Mpc}$ \\
      $\mathrm{M_{virial}}$
      & $2.1 \times 10^{15} \: \mathrm{M_{\odot}}$
      & $1.5 \times 10^{15} \: \mathrm{M_{\odot}}$ \\
      $\mathrm{M_{dm}}$
      & $2.0 \times 10^{15} \: \mathrm{M_{\odot}}$
      & $1.4 \times 10^{15} \: \mathrm{M_{\odot}}$ \\
      $\mathrm{M_{gas}}$
      & $1.2 \times 10^{14} \: \mathrm{M_{\odot}}$
      & $9.0 \times 10^{13} \: \mathrm{M_{\odot}}$ \\
      $\mathrm{L_{x}}$
      & $3.0 \times 10^{45} \: \mathrm{erg} \, \mathrm{s^{-1}}$
      & $5.0 \times 10^{44} \: \mathrm{erg} \, \mathrm{s^{-1}}$ \\
      $\mathrm{\overline{T}}$
      & $8.0 \times 10^{7} \: \mathrm{K} \: (6.9 \, \mathrm{keV})$
      & $6.3 \times 10^{7} \: \mathrm{K} \: (5.4 \, \mathrm{keV})$ \\
      $\mathrm{\sigma}$
      & $2,300 \: \mathrm{km} \, \mathrm{s^{-1}}$
      & $1,900 \: \mathrm{km} \, \mathrm{s^{-1}}$ \\
      $\mathrm{f_{cool}}$
      & 0.085
      & 0.117 \\
      \enddata
\end{deluxetable}

\setlength{\tabcolsep}{0.02in} 
\begin{deluxetable}{ccccccccccccc}
   \tabletypesize{\small}
   \rotate
   \tablewidth{0pt}
   \tablenum{2}
   \tablecolumns{12}
   \tablecaption{Subcluster Properties \label{tab:subclusters}}
   \startdata
      \hline
      \hline
      $\mathrm{{d}} \; \left(\mathrm{{Mpc}} \right)$ &
      $\mathrm{R_{virial}} \; \left(\mathrm{Mpc}\right)$ &
      $\mathrm{M_{virial}} \; \left(\mathrm{M_{\odot}}\right)$ &
      $\mathrm{M_{dm}} \; \left( \mathrm{M_{\odot}}\right)$ &
      $\mathrm{M_{gas}} \; \left( \mathrm{M_{\odot}} \right)$ &
      $\mathrm{L_{X}} \; \left(\mathrm{erg} \; \mathrm{s}^{-1}\right)$ &
      $\overline{\mathrm{T}} \; \left(\mathrm{K}\right)$ &
      $\mathrm{T_{core}} \; \left( K \right)$ &
      $\sigma \; \left(\mathrm{km} \; \mathrm{s}^{-1}\right)$ &
      $\mathrm{f_{cool}}$ &
      $\mathrm{t_{cool}} \; \left( \mathrm{yr} \right)$ &
      $\mathrm{Virial Ratio}$ \\
      \hline
      %%%
      2.2 &
      0.90 &
      $8.2 \times 10^{13}$ &
      $7.9 \times 10^{13}$ &
      $3.1 \times 10^{12}$ &
      $3.8 \times 10^{41}$ &
      $5.0 \times 10^{7}$ &
      $2.3 \times 10^{5}$ &
      820 &
      0.45 &
      $1.4 \times 10^{8}$ &
      $1.9^{\dag}$ \\
      %%%
      3.0 &
      0.49 &
      $1.4 \times 10^{13}$ &
      $1.3 \times 10^{13}$ &
      $8.1 \times 10^{11}$ &
      $1.6 \times 10^{40}$ &
      $9.4 \times 10^{6}$ &
      $8.2 \times 10^{3}$ &
      470 &
      0.61 &
      $2.1 \times 10^{8}$ &
      1.8\\
      %%%
      3.4 &
      0.41 &
      $7.8 \times 10^{12}$ &
      $7.2 \times 10^{12}$ &
      $5.4 \times 10^{11}$ &
      $2.5 \times 10^{40}$ &
      $9.4 \times 10^{6}$ &
      $1.0 \times 10^{5}$ &
      480 &
      0.81 &
      $7.7 \times 10^{8}$ &
      $2.2^{\dag}$ \\
      %%%
      4.6 &
      0.40 &
      $7.4 \times 10^{12}$ &
      $7.0 \times 10^{12}$ &
      $4.3 \times 10^{11}$ &
      $1.3 \times 10^{39}$ &
      $2.3 \times 10^{6}$ &
      $8.5 \times 10^{3}$ &
      340 &
      0.64 &
      $3.6 \times 10^{8}$ &
      1.4 \\
      %%%
      2.7 &
      0.36 &
      $5.4 \times 10^{12}$ &
      $5.4 \times 10^{12}$ &
      $7.5 \times 10^{10}$ &
      $4.8 \times 10^{38}$ &
      $1.8 \times 10^{7}$ &
      $8.4 \times 10^{3}$ &
      500 &
      0.46 &
      $1.1 \times 10^{9}$ &
      $4.1^{\dag}$ \\
      %%%
      3.9 &
      0.36 &
      $5.2 \times 10^{12}$ &
      $4.9 \times 10^{12}$ &
      $3.4 \times 10^{11}$ &
      $1.1 \times 10^{39}$ &
      $6.7 \times 10^{6}$ &
      $7.9 \times 10^{3}$ &
      360 &
      0.45 &
      $3.6 \times 10^{8}$ &
      1.7 \\
      %%%
      4.8 &
      0.32 &
      $3.8 \times 10^{12}$ &
      $3.7 \times 10^{12}$ &
      $8.3 \times 10^{10}$ &
      $3.3 \times 10^{38}$ &
      $5.5 \times 10^{6}$ &
      $8.9 \times 10^{3}$ &
      300 &
      0.47 &
      $1.0 \times 10^{9}$ &
      1.4 \\
      %%%
      2.5 &
      0.32 &
      $3.7 \times 10^{12}$ &
      $3.5 \times 10^{12}$ &
      $1.4 \times 10^{11}$ &
      $3.9 \times 10^{39}$ &
      $2.3 \times 10^{7}$ &
      $2.1 \times 10^{4}$ &
      310 &
      0.39 &
      $8.7 \times 10^{8}$ &
      1.6 \\
      %%%
      4.9 & 
      0.31 &
      $3.3 \times 10^{12}$ &
      $3.1 \times 10^{12}$ &
      $1.6 \times 10^{11}$ &
      $2.3 \times 10^{37}$ &
      $2.6 \times 10^{6}$ &
      $9.4 \times 10^{3}$ &
      260 &
      0.86 &
      $3.6 \times 10^{8}$ &
      1.2 \\
      %%%
      3.9 &
      0.28 &
      $2.5 \times 10^{12}$ &
      $2.4 \times 10^{12}$ &
      $1.0 \times 10^{11}$ &
      $4.4 \times 10^{36}$ &
      $2.6 \times 10^{6}$ &
      $9.4 \times 10^{3}$ &
      260 &
      0.86 &
      $3.6 \times 10^{8}$ &
      1.2 \\
      %%%
      4.1 &
      0.27 &
      $2.1 \times 10^{12}$ &
      $2.0 \times 10^{12}$ &
      $1.3 \times 10^{11}$ &
      $1.8 \times 10^{37}$ &
      $2.7 \times 10^{6}$ &
      $7.8 \times 10^{3}$ &
      200 &
      0.81 &
      $5.5 \times 10^{8}$ &
      0.9 \\
      %%%
      4.9 &
      0.26 &
      $1.9 \times 10^{12}$ &
      $1.8 \times 10^{12}$ &
      $1.7 \times 10^{11}$ &
      $2.2 \times 10^{37}$ &
      $1.4 \times 10^{6}$ &
      $8.6 \times 10^{3}$ &
      200 &
      0.82 &
      $4.3 \times 10^{8}$ &
      0.9 \\
      %%%
      4.7 &
      0.22 &
      $1.2 \times 10^{12}$ &
      $1.1 \times 10^{12}$ &
      $7.0 \times 10^{10}$ &
      $4.5 \times 10^{36}$ &
      $7.7 \times 10^{5}$ &
      $8.7 \times 10^{3}$ &
      130 &
      0.76 &
      $7.7 \times 10^{8}$ &
      0.8 \\
      %%%
      %2.8 &
      %0.21 &
      %$1.0 \times 10^{12}$ &
      %$9.7 \times 10^{11}$ &
      %$8.5 \times 10^{10}$ &
      %$1.0 \times 10^{39}$ &
      %$1.0 \times 10^{7}$ &
      %$2.7 \times 10^{4}$ &
      %440 &
      %0.57 &
      %$1.9 \times 10^{9}$ &
      %6.5 \\
      %%%
      %3.5 &
      %0.21 &
      %$1.0 \times 10^{12}$ &
      %$9.7 \times 10^{11}$ &
      %$3.5 \times 10^{10}$ &
      %$1.5 \times 10^{37}$ &
      %$3.8 \times 10^{6}$ &
      %$9.8 \times 10^{3}$ &
      %260 &
      %0.64 &
      %$1.6 \times 10^{9}$ &
      %2.1 \\
      %%%
      %3.1 &
      %0.17 &
      %$5.6 \times 10^{11}$ &
      %$5.1 \times 10^{11}$ &
      %$5.6 \times 10^{10}$ &
      %$3.8 \times 10^{37}$ &
      %$6.0 \times 10^{5}$ &
      %$1.0 \times 10^{4}$ &
      %160 &
      %0.41 &
      %$5.1 \times 10^{9}$ &
      %2.4 \\
      \enddata
      \tablecomments{\dag \ Subcluster passed within 2 Mpc of the main cluster}
\end{deluxetable}

\clearpage

\clearpage

\clearpage

\clearpage

\clearpage

\clearpage

\begin{figure}
\plotone{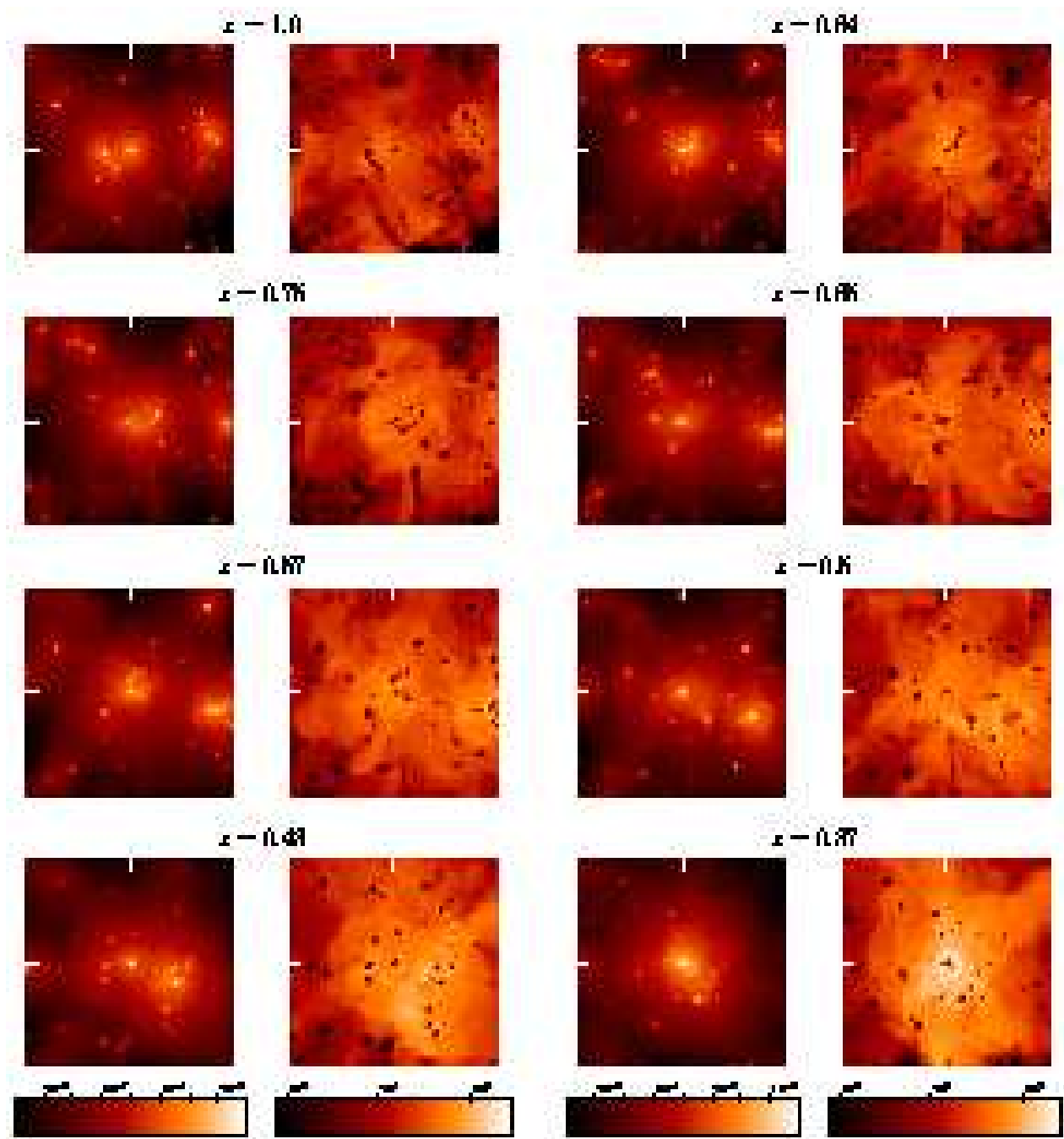}
\figcaption[f7a.eps]{Projected, normalized X-ray surface brightness (left)
  and temperature maps (right) for C1 from a redshift of 1 to the present
  epoch in intervals of approximately 500 million years. The images are 5
  $h^{-1}$ Mpc on a side at the present epoch.  All images were prepared
  on the same scale so that a given color corresponds to the same value
  throughout the table of images.
  The temperature is in Kelvin and
  the tick marks to the left and above each image mark the cluster center
  of mass.}
\end{figure}

\clearpage

\begin{figure}
\plotone{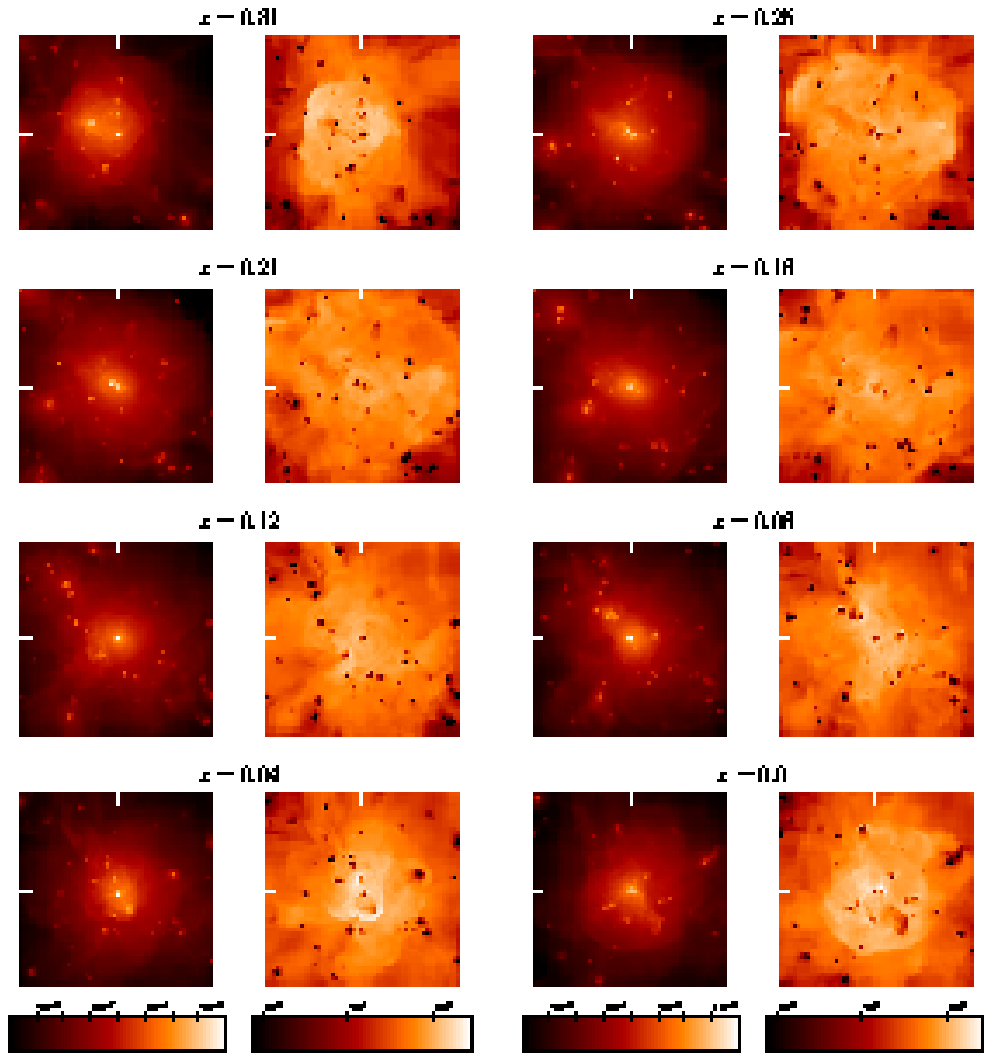}
\figurenum{7}
\figcaption[f7b.eps]{Projected, normalized X-ray surface brightness (left)
  and temperature maps (right) for C1 from a redshift of 1 to the present
  epoch in intervals of approximately 500 million years.  As before the images
  are 5 $h^{-1}$ Mpc on a side.  All images were prepared on the same scale
  so that a given color corresponds to the same value
  throughout the table of images and the range of values are depicted in
  the color bars at bottom.\label{fig:cl00_table}}
\end{figure}

\clearpage

\begin{figure}
\plotone{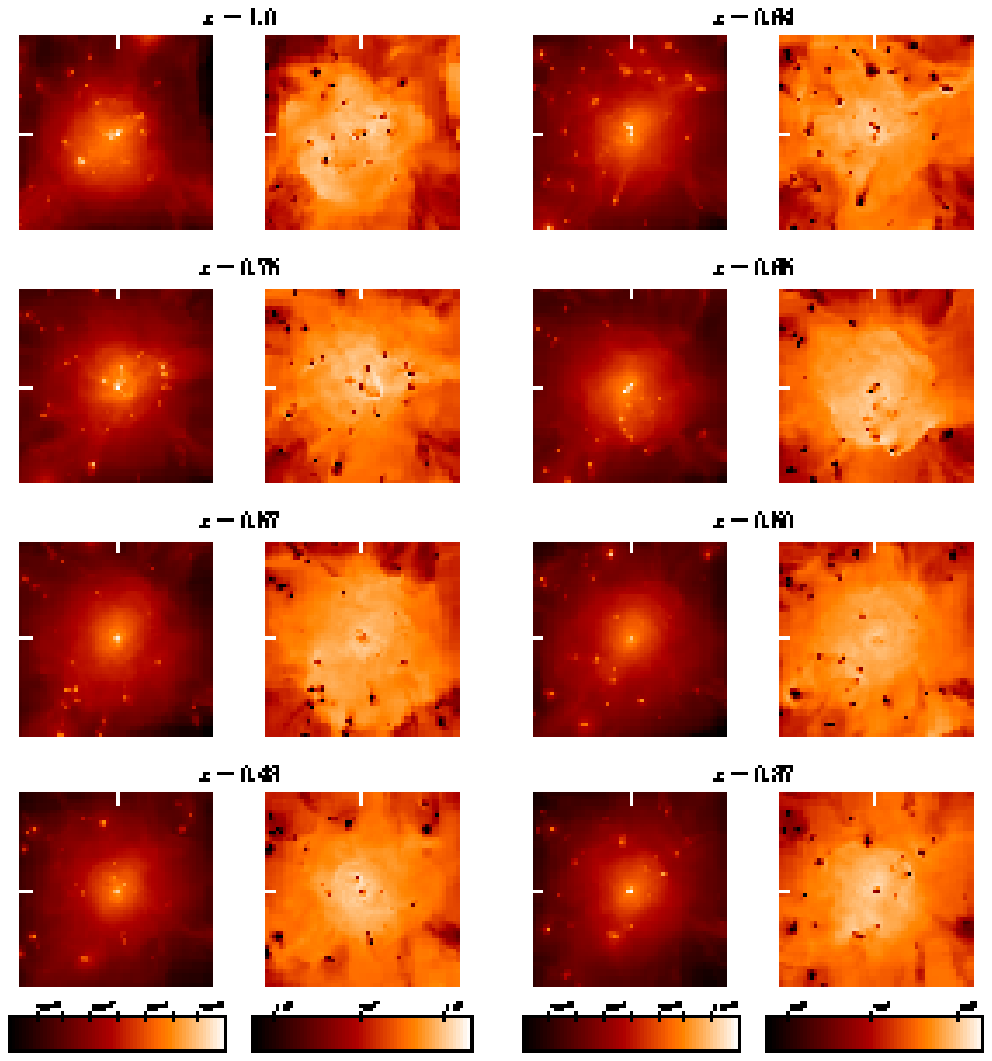}
\figcaption[f8a.eps]{Projected, normalized X-ray surface brightness (left)
  and temperature maps (right) for C2 from a redshift of 1 to the present 
  epoch in intervals of approximately 500 million years.  As before the images 
  are 5 $h^{-1}$ Mpc on a side.  All images were prepared on the same scale so 
  that a given color corresponds to the same intensity or temperature level 
  throughout the table of images.}
\end{figure}

\clearpage

\begin{figure}
\plotone{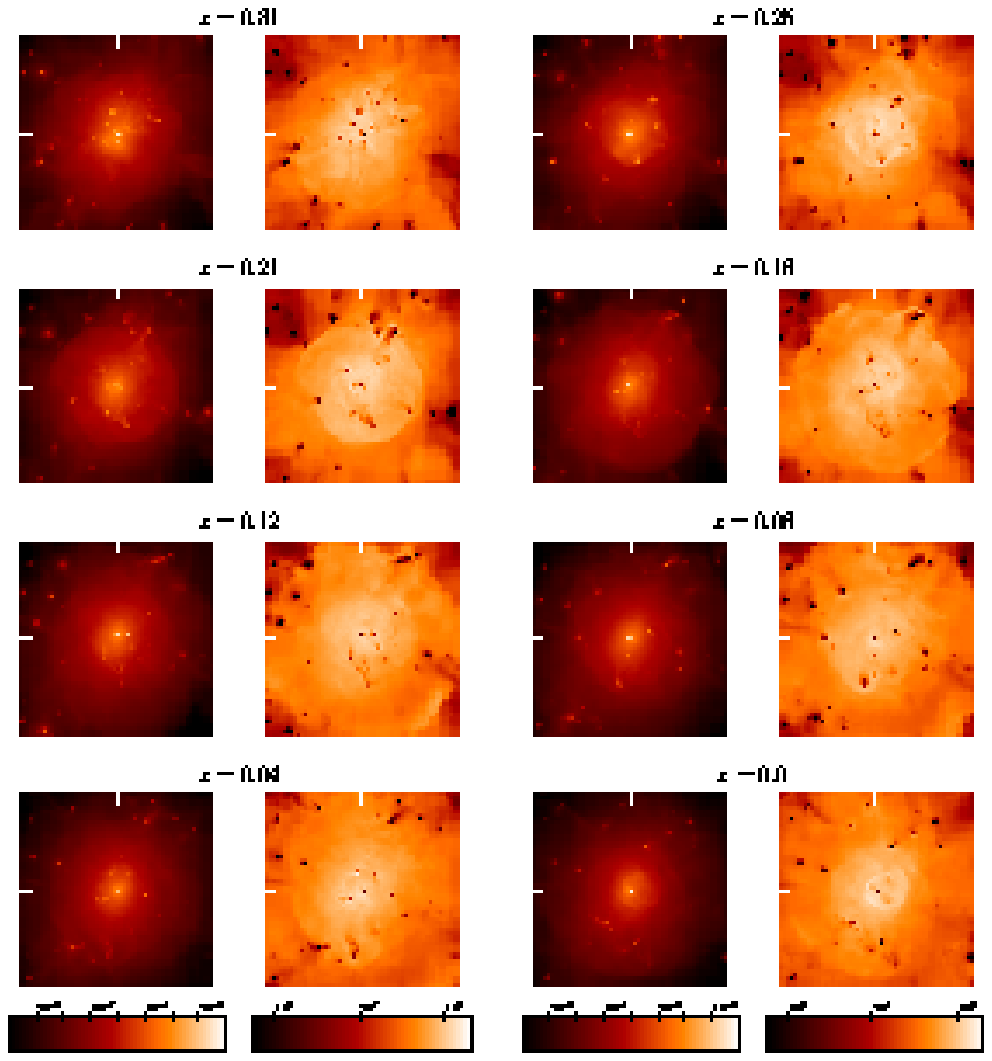}
\figurenum{8}
\figcaption[f8b.eps]{Projected, normalized X-ray surface brightness (left)
  and temperature maps (right) for Cluster 2 from a redshift of 1 to the
  present epoch in intervals of approximately 500 million years.  As before
  the images are 5 $h^{-1}$ Mpc on a side.  All images were prepared on the
  same scale so that a given color corresponds to the same surface brightness
  or temperature level throughout the table of images.  \label{fig:cl01_table}}
\end{figure}

\clearpage

\begin{figure}
\plotone{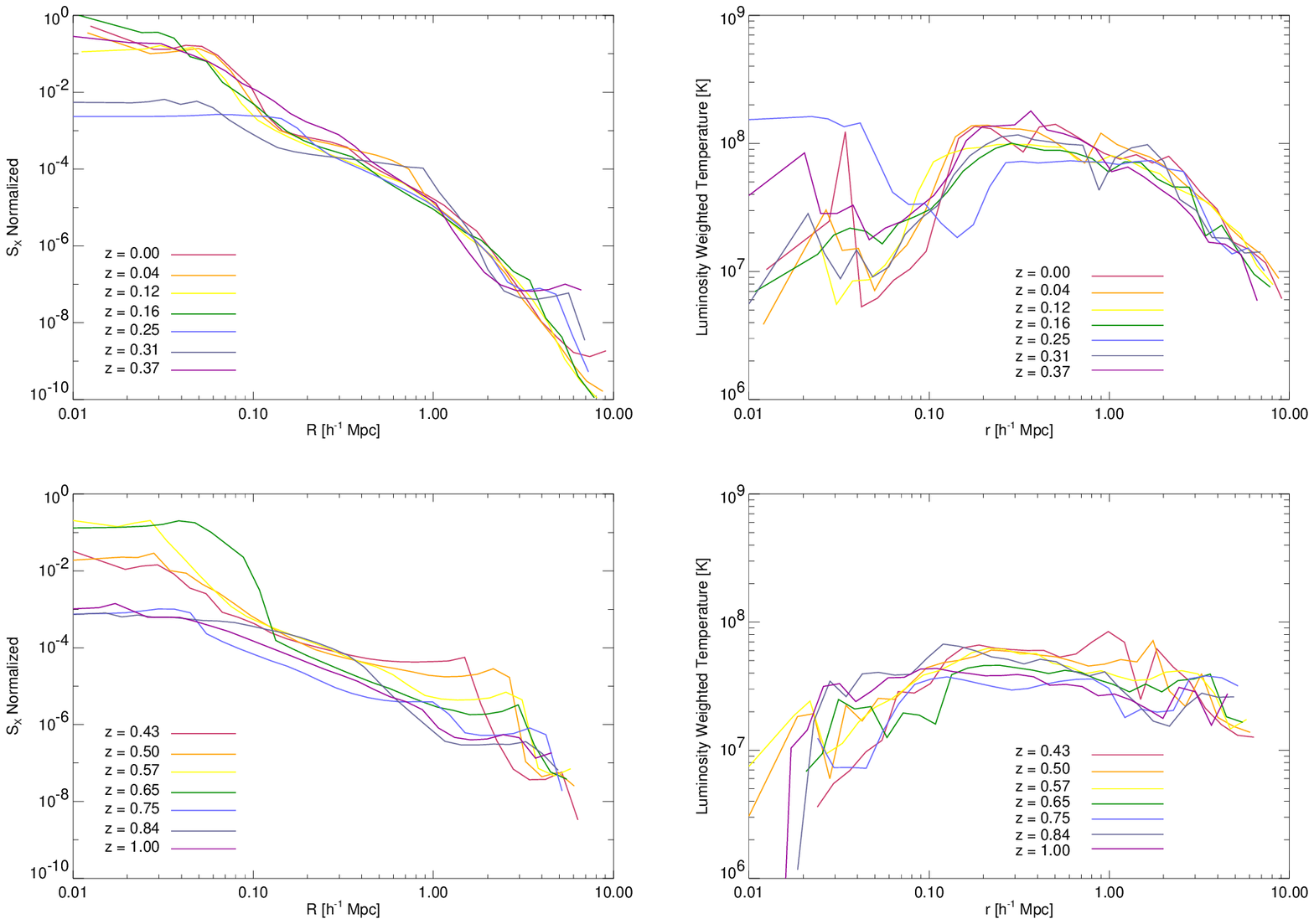}
\figcaption[f9.eps]{X-ray surface brightness profiles in the 1 to 10 keV band, 
  normalized to the maximum value and spherically-averaged, luminosity-weighted 
  temperature profiles for C1 at the indicated redshifts.  
  \label{fig:cl00_prof_table}}
\end{figure}

\clearpage

\begin{figure}
\plotone{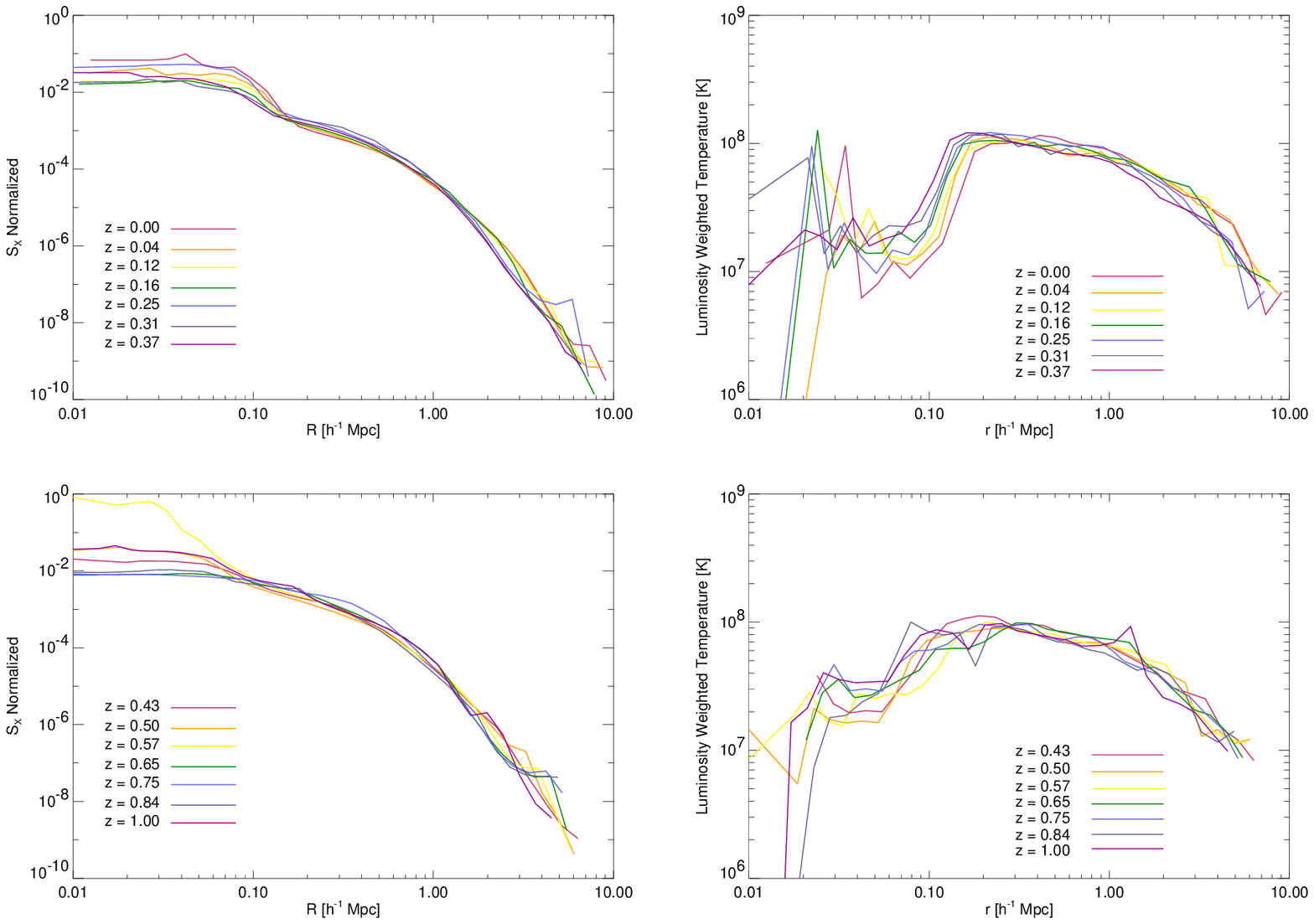}
\figcaption[f10.eps]{X-ray surface brightness profiles in the  1 to 10 keV band, 
  normalized to the maximum value and spherically-averaged, emission-weighted 
  temperature profiles for C2 at the indicated redshifts.  
  \label{fig:cl01_prof_table}}
\end{figure}

\begin{figure}
\plotone{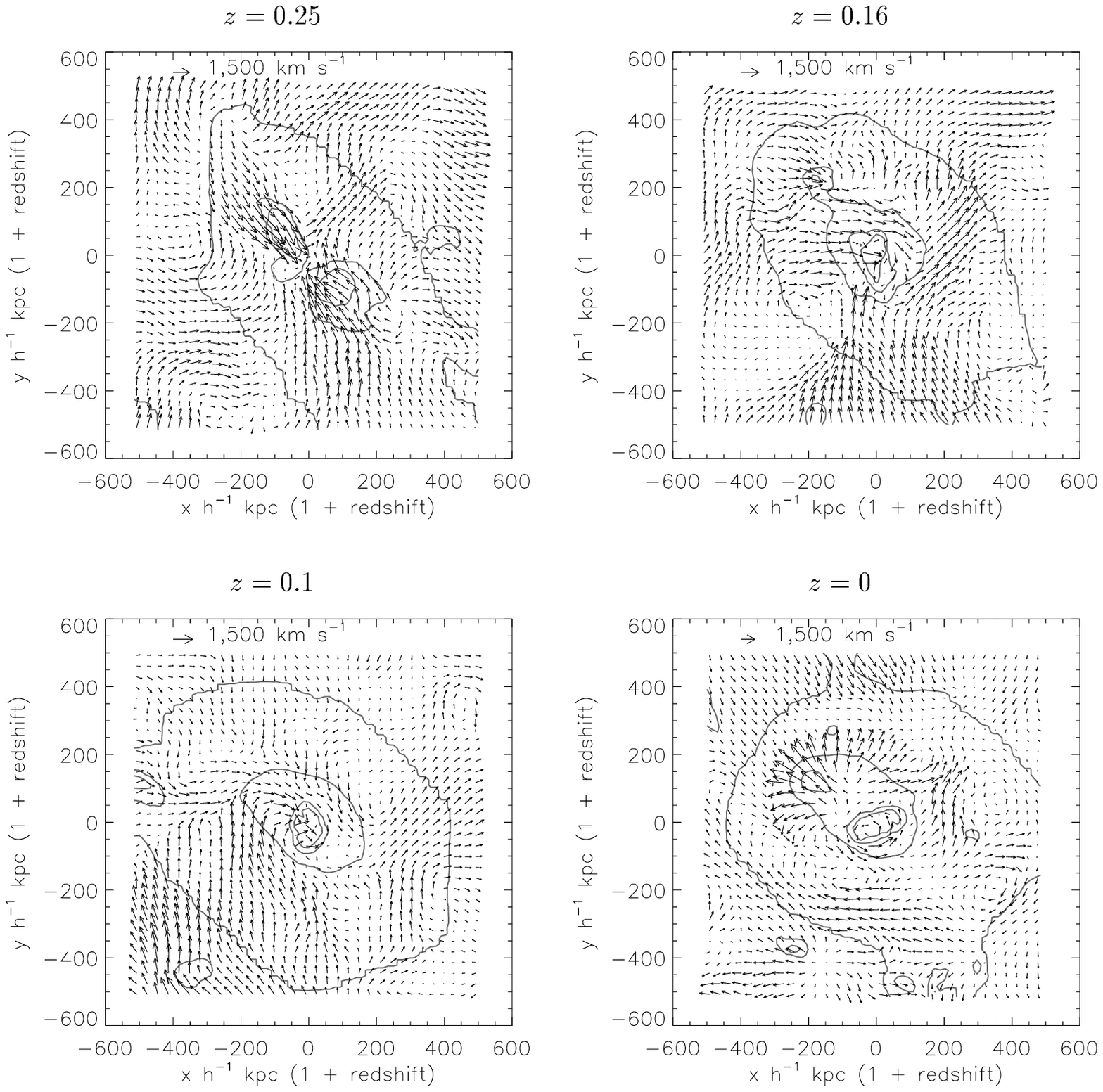}
\figcaption[f11.eps]{Velocity fields in the x-y plane for cluster C1 at the 
  indicated epochs.  The scale of the velocity vectors is indicated by the 
  vector with magnitude 1,500 $\mathrm{km} \; \mathrm{s}^{-1}$ at the top 
  of each plot.  Projected X-ray surface brightness contours are also shown 
  to help outline the flow field.  The four contours are spaced logarithmically 
  from $1 \times 10^{-2}$ to $1 \times 10^{-5}$ of the peak value.  
  \label{fig:cl00_vel}}
\end{figure}

\begin{figure}
\begin{center}
\includegraphics[scale=0.83]{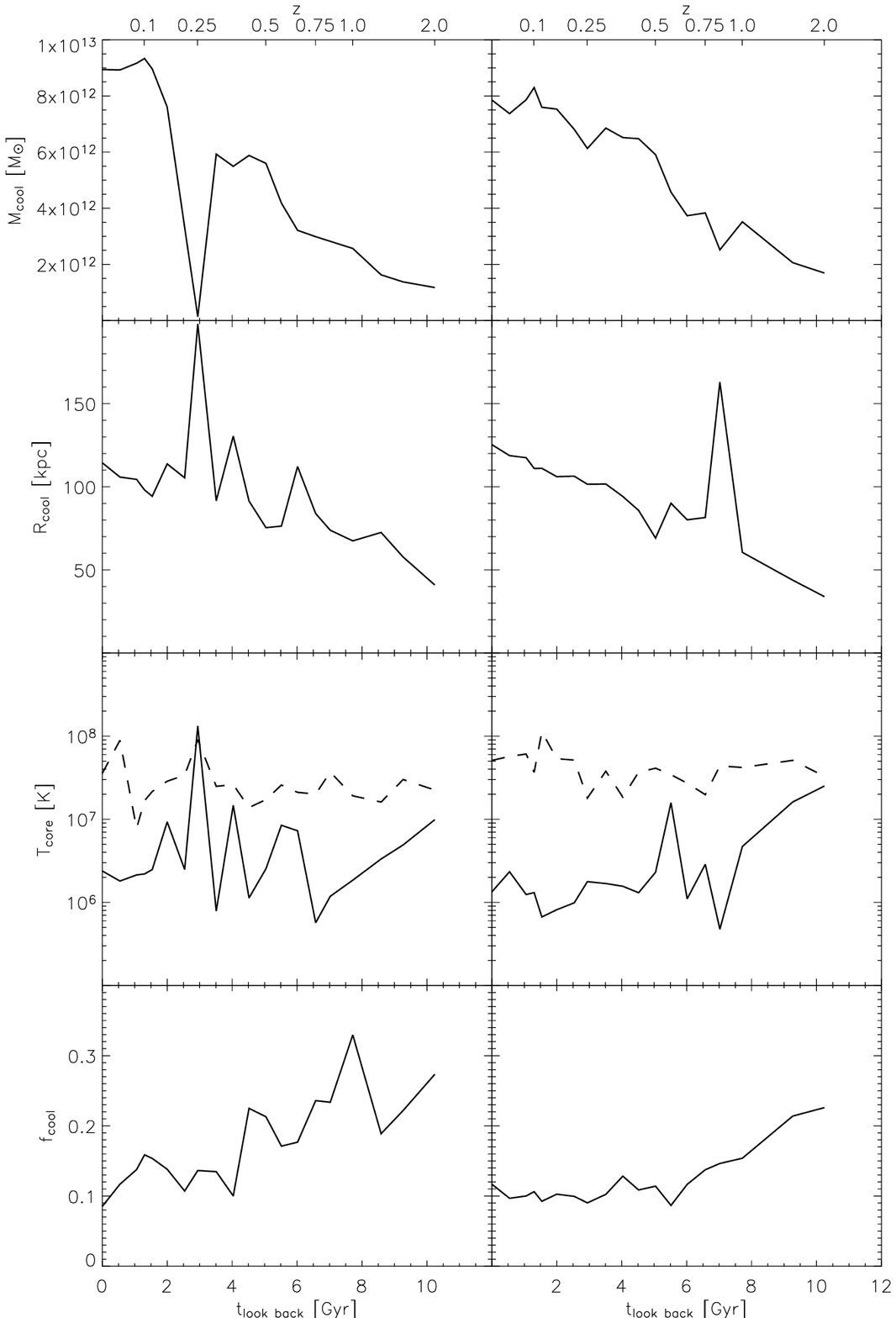}
\figcaption[f12.eps]{Mass of cold gas within 100 $\mathrm{h}^{-1}$ kpc of
  the cluster center (top row), the cooling radius (second row), central
  core temperature (third row), and fraction of cool gas (bottom row)
  all as a function of lookback time for the
  simulated clusters C1 (left) and C2 (right). The core temperature is
  measured in two ways; the solid curves correspond to the physical gas
  temperature averaged over a sphere of radius 50 $\mathrm{h}^{-1}$ kpc
  centered on the cluster center of mass while the dashed curves correspond
  to the average within a 50 $\mathrm{h}^{-1}$ kpc circle of the projected,
  emission-weighted temperature field.  The redshifts for a given lookback
  time are shown at top for reference.
  \label{fig:cool_parameters}}
  \end{center}
\end{figure}

\begin{figure}
\plotone{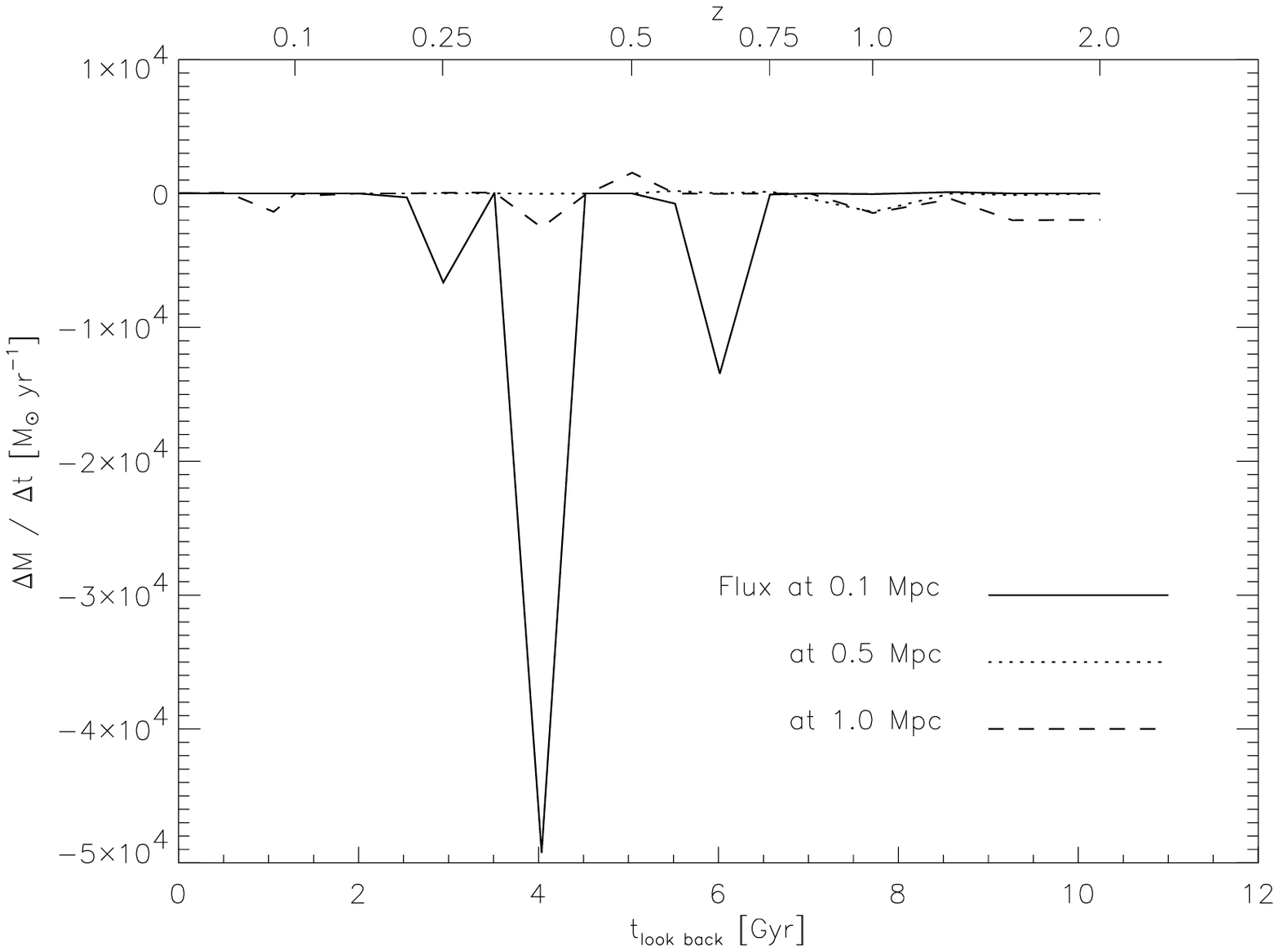}
\figcaption[f13.eps]{Flux of cool gas ( in $\mathrm{M}_{\odot} / \mathrm{yr}$ ) 
  through spheres of radius 0.1, 0.5, and 1.0 Mpc centered on the cluster center 
  of mass throughout the C1 simulation. \label{fig:clrc00_cold_flux}}
\end{figure}

\begin{figure}
\plotone{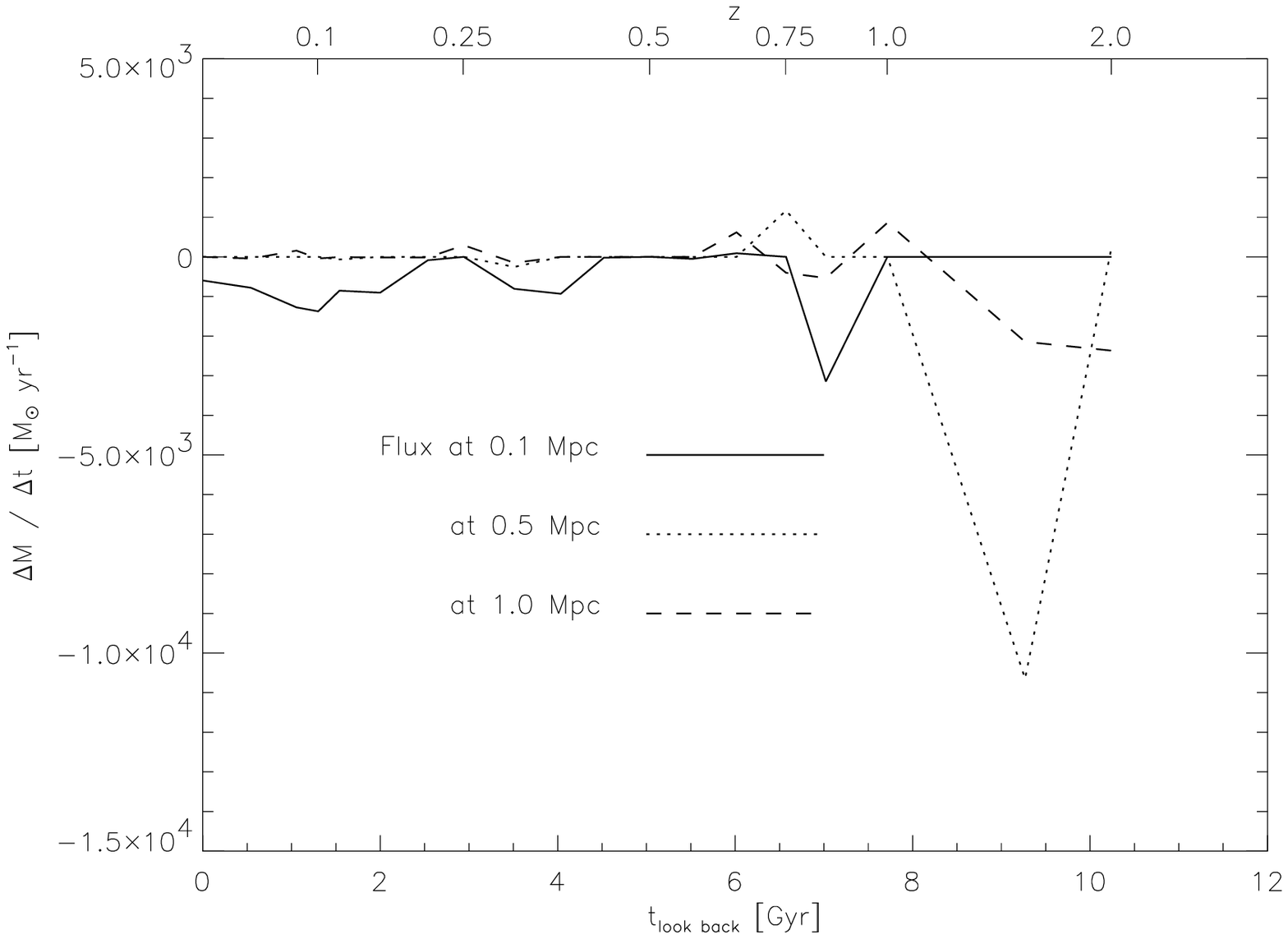}
\figcaption[f14.eps]{Flux of cool gas ( in $\mathrm{M}_{\odot} / \mathrm{yr}$ ) 
  through spheres of radius 0.1, 0.5, and 1.0 Mpc centered on the cluster center 
  of mass for the C2 simulation. \label{fig:clrc01_cold_flux}}
\end{figure}

\begin{figure}
\plotone{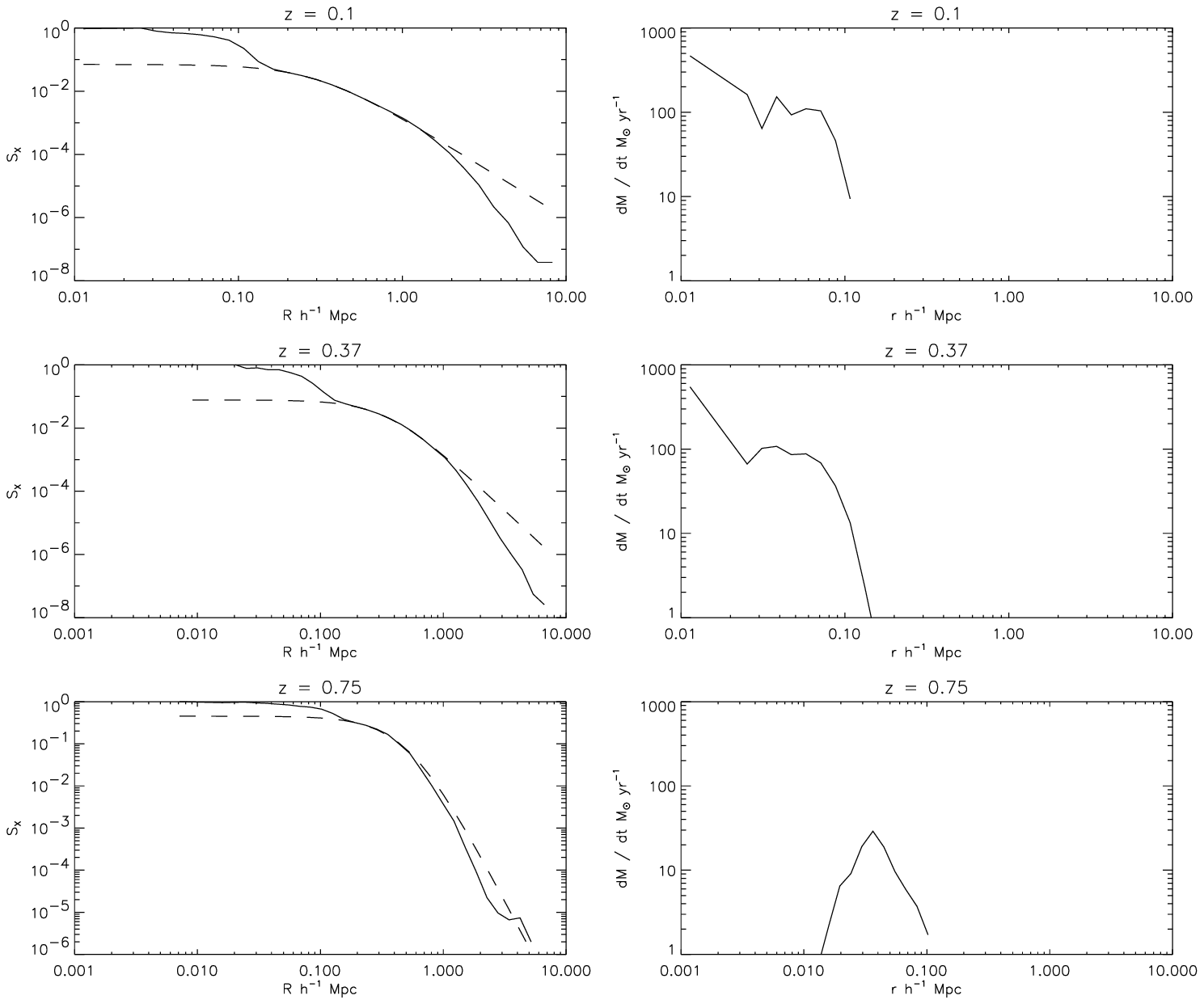}
\figcaption[f15.eps]{In the left column the predicted surface brightness profile
  (solid curve) and the fit isothermal profile (dashed line) for cluster C2 at 
  the indicated redshift.  The right column shows the implied cooling flow rate 
  given the temperature profile. At a redshift of $z = 0.75$ the cluster is in 
  the final stages of a major merger and the presence of two luminosity maxima 
  on opposite sides of the cluster center of mass gives rise to the decline in 
  the inferred cooling rate.\label{fig:c2_mdot_excess}}
\end{figure}

\begin{figure}
\plotone{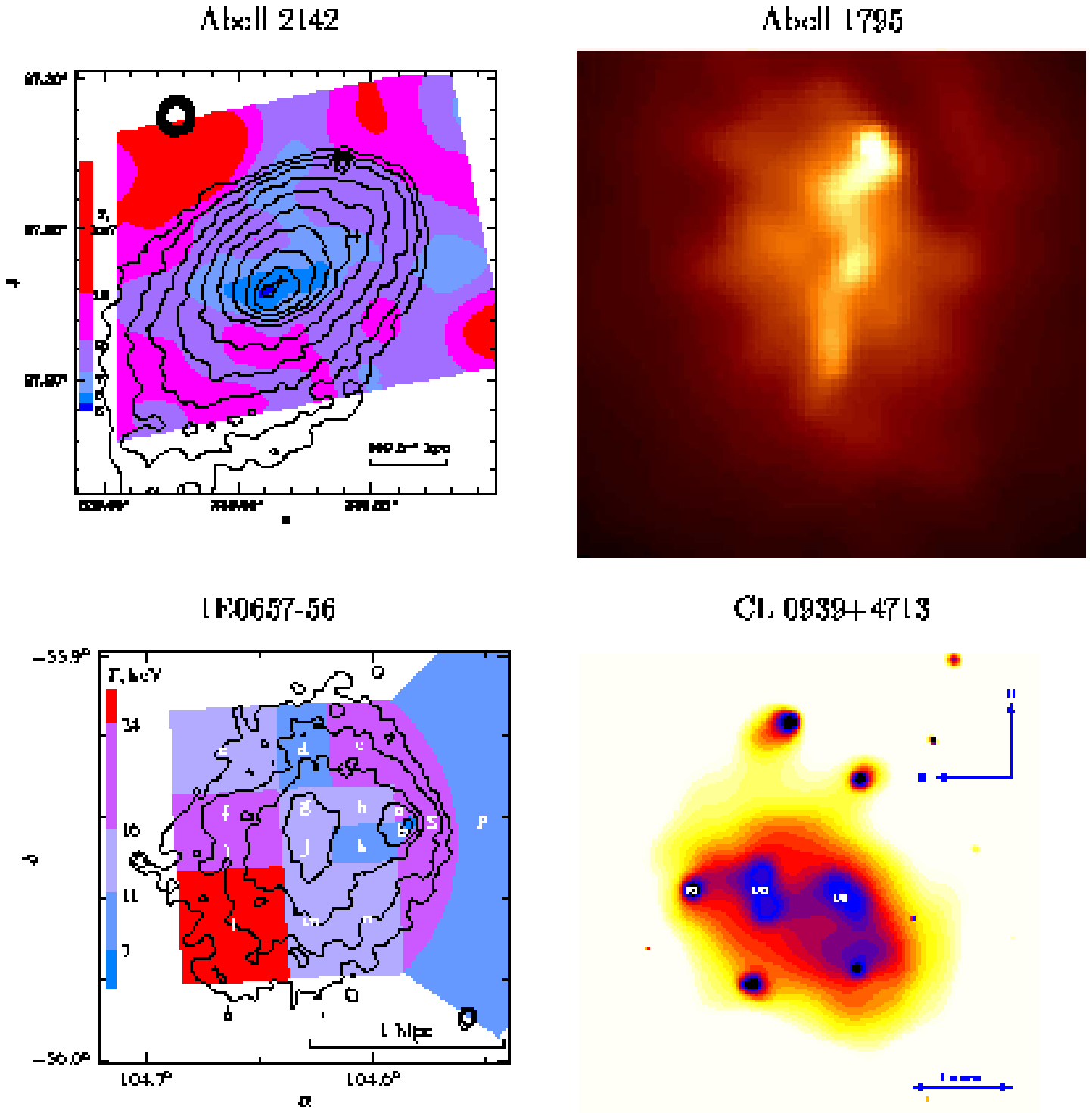}
\figcaption[f16.eps]{ Recent results from \textit{Chandra} and \textit{XMM} 
  emphasizing the rich and complex nature of cool core clusters.  Clockwise 
  from upper left: Abell 2142 with its irregular temperature  distribution and  
  ``cool fronts'' \citep{markevitch_2142}, the filamentary distribution of cool 
  material in the core of Abell 1795 \citep{fabian01}, the colliding ``bullet 
  subcluster'' system 1E0657-56 \citep{markevitch02}, and finally a more distant 
  cluster with significant central substructure, CL 0939+4713 (De Filippis 
  \textit{et al.} 2001).  \label{fig:obs_comparisson}}
\end{figure}


\begin{thebibliography}

\bibitem[Bahcall \& Cen (1993)]{bahcall93} Bahcall, N.~A., \& Cen, R. 1993,
         \apj, 407, L49

\bibitem[Balogh \textit{et al.}~(2001)]{balogh01} Balogh, M.~L., Pearce, F.~R.,
         Bower, R.~G., \& Kay, S.~T. 2001, \mnras, 326, 1228

\bibitem[Bryan, Abel \& Norman (2001)]{bryan01} Bryan, G.~L., Abel, T., \&
         Norman, M.~L. 2001 in Proceedings of Supercomputing 2001
         http://www.sc2001.org/

\bibitem[Burns \textit{et al}.~(1997)]{burns97} Burns, J.~O., Loken, C.,
         G\'{o}mez, P., Rizza, E., Bliton, M., \& Ledlow, M. 1997 in Galactic
         and Cluster Cooling Flows, ed. N. Soker, (San Francisco: ASP), 115, p21

\bibitem[Chandran \& Cowley (1998)]{chandran98} Chandran, B.~D.~G., \& Cowley,
         S.~C. 1998, \prl, 80, 3077

\bibitem[Cen \& Ostriker (1992)]{cen92} Cen, R., \& Ostriker, J.~P. 1992, \apj,
         393, 22

\bibitem[Cen \& Ostriker (1999)]{cen99} Cen, R., \& Ostriker, J.~P. 1999, \apjl,
         519, L109

\bibitem[Colella \& Woodward (1984)]{colella84} Colella, P., \& Woodward, P.~R. 1984,
         J. Comput. Phys., 54, 174

\bibitem[Cowie \& Binney (1977)]{cowie77} Cowie, L.~L., \& Binney, J. 1977, \apj,
         215, 723

\bibitem[Dav\'{e} \textit{et al.}~(2001)]{dave01} Dav\'{e} R., Cen, R., Ostriker,
         J.~P., Bryan, G.~L., Hernquist, L., Katz, N., Weinberg, D.~H., Norman,
         M.~L. \& O'Shea, B. 2001, \apj, 552, 473

\bibitem[De Filippis \textit{et al.}~(2001)]{filippis01} De Filippis, E., Schindler,
S., \& Castillo-Morales, A. 2001 in New Visions of the X-Ray Universe in the XMM-Newton
and Chandra Era, eds. F. Jansen \textit{et al.} ESA WPP Conference Series 488
(astro-ph/0201349)

\bibitem[Evrard \textit{et al.}~(1994)]{evrard94} Evrard, A.~E., Summers, F.~J.,
         \& Davis, M. 1994, \apj, 422, 11

\bibitem[Fabian \& Nulsen (1977)]{fabian77} Fabian, A.~C., \& Nulsen, P.~E.~J. 1977,
         \mnras, 180, 479

\bibitem[Fabian \& Daines (1991)]{fabian91} Fabian, A.~C., \& Daines, S.~J. 1991,
         \mnras, 252, 17

\bibitem[Fabian (1994)]{fabian-araa-94} Fabian, A.~C. 1994, \araa, 32, 277

\bibitem[Fabian \textit{et al.}~(1994)]{fabian94} Fabian, A.~C., Canizares, C.~R.,
         \& B\"{o}hringer H. 1994, \apj, 425, 40

\bibitem[Fabian \textit{et al.}~(2001)]{fabian01} Fabian, A.~C., Sanders, J.~S.,
         Ettori, S., Taylor, G.~B., Allen, S.W., Crawford, C.~S., Iwasawa, K.,
         \& Johnstone, R.~M. 2001, \mnras, 321, L33.

\bibitem[Fabian (2002)]{fabian02} Fabian, A.~C. 2002 in Lighthouses of the Universe,
         eds. Gilfanov, M., Sunyaev, R., \& E. Churazov (Heidelberg: Springer-Verlag)
         (astro-ph/0201386)

\bibitem[Fabian, Voigt, & Morris (2002)]{faiban-cond} Fabian, A.~C., Voigt, L.~M.,
        \& Morris, R.~G. 2002, \mnras, 335, L71

\bibitem[Frenk \textit{et al.}~(1996)]{frenk96} Frenk, C.~S., Evrard, A.~E.,
         White, S.~D.~M., Summers, F.~J. 1996, \apj, 472, 460

\bibitem[G\'{o}mez \textit{et al.}~(2002)]{gomez02} G\'{o}mez, P.~L., Loken, C.,
         Roettiger, K., \& Burns, J.~O. 2002, \apj, 569, 122

\bibitem[Katz \& White (1993)]{katz93} Katz, N., \& White, S.~D.~M. 1993, \apj,
         412, 455

\bibitem[Kempner \textit{et al.}~(2002)]{kempner02} Kempner, J., Sarazin, C.~L.,
         \& Ricker, P.~R. 2002, \apj, in press

\bibitem[Lewis \textit{et al.}~(2000)]{lewis00} Lewis, G.~F., Babul, A., Katz, N.,
         Quinn, T., Hernquist, L., \& Weinberg, D.~H. 2000, \apj, 536, 623

\bibitem[Loken \textit{et al.}~(1999)]{loken99} Loken, C., Melott, A.~L., \&
         Miller, C.~J. 1999, \apjl, 520, L5

\bibitem[Loken \textit{et al.}~(2002)]{loken02} Loken, C., Norman, M.~L., Nelson, E.,
         Burns, J.~O., Bryan, G.~L., \& Motl, P.~M. 2002, \apj, 579, 571

\bibitem[Markevitch \textit{et al.}~(2000)]{markevitch_2142} Markevitch, M. \textit{et al.} 2000,
         \apj, 541, 542

\bibitem[Markevitch \textit{et al.}~(2001)]{markevitch_2163} Markevitch, M., Vikhlinin,
         A., Mazzotta, P., \& Van~Speybroeck L. 2001 in X-ray Astronomy 2000, eds.
         R.~Giacconi, L.~Stella, \& S.~Serio (San Francisco: ASP)

\bibitem[Markevitch \textit{et al.}~(2002)]{markevitch02} Markevitch, M., Gonzalez,
         A.~H., David, L., Vikhlinin, A., Murray, S., Forman, W.~R., Jones, C., \&
         Tucker, W. 2002, \apjl, 567, L27

\bibitem[Mazzotta \textit{et al.}~(2001)]{mazzotta01} Mazzotta, P., Markevitch, M.,
         Vikhlinin, A., Forman, W.~R., David, L.~P., \& Van~Speybroeck, L. 2001, \apj,
         555, 205

\bibitem[Mazzotta \textit{et al.}~(2002)]{mazzotta02} Mazzotta, P., Markevitch, M.,
         Forman, W.~R., Jones, C., Vikhlinin, A., \& Van~Speybroeck, L. 2002, \apj
          in press

\bibitem[McGlynn \& Fabian (1984)]{mcglynn84} McGlynn, T.~A., \& Fabian, A.~C. 1984,
         \mnras, 208, 709

\bibitem[Narayan \& Medvedev (2001)]{narayan01} Narayan, R., \& Medvedev, M.~V. 2001,
         \apjl, 562, L132

\bibitem[Norman \& Bryan (1999)]{norman99} Norman, M.~L., \& Bryan, G.~L. 1999
         in ASSL Vol. 240: Numerical Astrophysics, eds. S.~M.~Miyama, K.~Tomisaka,
         \& T.~Hanawa, (Boston: Kluwer), 19

\bibitem[Pearce \textit{et al.}~(1999)]{pearce99} Pearce, F.~R., Jenkins, A., Frenk,
          C.~S., Colberg, J.~M., White, S.~D.~M., Thomas, P.~A., Counchman, H.~M.~P.,
         Peacock, J.~A., \& Efsathiou, G. 1999, \apjl, 521, L99

\bibitem[Pearce \textit{et al.}~(2000)]{pearce00} Pearce, F.~R., Thomas, P.~A.,
         Couchman, H.~M.~P., \& Edge, A.~C. 2000, \mnras, 317, 1029

\bibitem[Peres \textit{et al.}~(1998)]{peres98} Peres, C.~B., Fabian, A.~C., Edge, A.~C.,
         Allen, S.~W., Johnstone, R.~M., \& White, D.~A. 1998, \mnras, 298, 416

\bibitem[Ricker \& Sarazin (2001)]{ricker01} Ricker, P.~M., \& Sarazin, C.~L. 2001,
         \apj, 561, 621

\bibitem[Ritchie \& Thomas (2002)]{ritchie02} Ritchie, B.~W., \& Thomas, P.~A. 2002,
         \mnras, 329, 675

\bibitem[Roettiger \textit{et al.}~(1996)]{roettiger96} Roettiger, K., Burns, J.~O.,
         \& Loken, C. 1996, \apj, 473, 651

\bibitem[Sanders \& Fabian (2002)]{sanders02} Sanders, J.~S., \& Fabian, A.~C. 2002,
         \mnras, 331, 273

\bibitem[Schuecker \textit{et al.}~(2001)]{schuecker01} Schuecker, P., B\"{o}hringer,
          H., Reiprich, T.~H., \& Feretti, L. 2001, \aap, 378, 408

\bibitem[Sparks (1992)]{sparks92} Sparks, W.~B. 1992, \apj, 399, 66

\bibitem[Suginohara \& Ostriker (1998)]{suginohara98} Suginohara, T., \& Ostriker, J.~P.
         1998, \apj, 507, 16

\bibitem[Thomas \& Couchman (1992)]{thomas92} Thomas, P.~A., \& Couchman, H.~M.~P. 1992,
         \mnras, 257, 11

\bibitem[Tittley \textit{et al.}~(2001)]{tittley01} Tittley, E.~R., Pearce, F.~R., \&
         Couchman, H.~M.~P. 2001, \apj, 561, 69

\bibitem[Valdarnini (2002)]{valdarnini02}Valdarnini, R. 2002, \apj, 567, 741

\bibitem[Vikhlinin \textit{et al.}~(2001a)]{vikhlinin01} Vikhlinin, A., Markevitch, M.,
         Forman, W.~R., \& Jones, C. 2001a, \apjl, L87

\bibitem[Vikhlinin \textit{et al.}~(2001b)]{vikhlinin_3667} Vikhlinin, A., Markevitch, M.,
         \& Murray, S.~S. 2001b, \apj, 551, 160

\bibitem[Westbury \& Henriksen (1992)]{westbury92} Westbury, C.~F., \& Henriksen, R.~N.
         1992 \apj, 338, 64

\bibitem[White \textit{et al.}~(1997)]{white97} White, D.~A., Jones, C., \& Forman, W.
         1997 \mnras, 292, 419

\bibitem[Zakamska \& Narayan (2003)]{zakamska03} Zakamska, N.~L., \& Narayan, R. 2003,
         \apj, 582, 162

\end{thebibliography}
\end{document}